\documentclass[twocolumn]{aastex63}
\usepackage{amsmath}

\shorttitle{Gas metallicity gradients in GASP and MaNGA galaxies}
\shortauthors{Franchetto et al.}

\begin{document}

\title{GASP and MaNGA surveys shed light on the enigma of the gas metallicity gradients in disk galaxies}

\correspondingauthor{Andrea Franchetto}
\email{andrea.franchetto@phd.unipd.it}

\author[0000-0001-9575-331X]{Andrea Franchetto}
\affiliation{Dipartimento di Fisica e Astronomia ``Galileo Galilei'', Universit\`a di Padova, vicolo dell'Osservatorio 3, IT-35122, Padova, Italy}
\affiliation{INAF - Astronomical Observatory of Padova, vicolo dell'Osservatorio 5, IT-35122 Padova, Italy}

\author[0000-0003-2589-762X]{Matilde Mingozzi}
\affiliation{Space Telescope Science Institute, 3700 San Martin Drive, Baltimore, MD 21218, USA}

\author[0000-0001-8751-8360]{Bianca M. Poggianti}
\affiliation{INAF - Astronomical Observatory of Padova, vicolo dell'Osservatorio 5, IT-35122 Padova, Italy}

\author[0000-0003-0980-1499]{Benedetta Vulcani}
\affiliation{INAF - Astronomical Observatory of Padova, vicolo dell'Osservatorio 5, IT-35122 Padova, Italy}

\author[0000-0002-8372-3428]{Cecilia Bacchini}
\affiliation{INAF - Astronomical Observatory of Padova, vicolo dell'Osservatorio 5, IT-35122 Padova, Italy}

\author[0000-0002-7296-9780]{Marco Gullieuszik}
\affiliation{INAF - Astronomical Observatory of Padova, vicolo dell'Osservatorio 5, IT-35122 Padova, Italy}

\author[0000-0002-1688-482X]{Alessia Moretti}
\affiliation{INAF - Astronomical Observatory of Padova, vicolo dell'Osservatorio 5, IT-35122 Padova, Italy}

\author[0000-0002-8238-9210]{Neven Tomi\v{c}i\'c}
\affiliation{INAF - Astronomical Observatory of Padova, vicolo dell'Osservatorio 5, IT-35122 Padova, Italy}

\author[0000-0002-7042-1965]{Jacopo Fritz}
\affiliation{Instituto de Radioastronom\'ia y Astrof\'isica, UNAM, Campus Morelia, A.P. 3-72, C.P. 58089, M\'exico}


\begin{abstract}

Making use of both MUSE observations of 85 galaxies from the survey GASP (GAs Stripping Phenomena in galaxies with MUSE) and a large sample from MaNGA (Mapping Nearby Galaxies at Apache Point Observatory survey) we investigate the distribution of gas metallicity gradients as a function of stellar mass, for local cluster and field galaxies. Overall, metallicity profiles steepen with increasing stellar mass up to $10^{10.3}\,{\rm M_\odot}$ and flatten out at higher masses. Combining the results from the metallicity profiles and the stellar mass surface density gradients, we propose that the observed steepening is a consequence of local metal enrichment due to in-situ star formation during the inside-out formation of disk galaxies. The metallicity gradient-stellar mass relation is characterized by a rather large scatter, especially for  $10^{9.8}<{\rm M_\ast/M_\odot}<10^{10.5}$, and we demonstrate that metallicity gradients anti-correlate with the galaxy gas fraction.
Focusing on the galaxy environment, at any given stellar mass, cluster galaxies have systematically flatter metallicity profiles than their field counterparts. Many sub-populations coexist in clusters: galaxies with shallower metallicity profiles appear to have fallen into their present host halo sooner and have experienced the environmental effects for a longer time than cluster galaxies with steeper metallicity profiles. Recent galaxy infallers, like galaxies currently undergoing ram-pressure stripping, show metallicity gradients more similar to those of field galaxies, suggesting they have not felt the effect of the cluster yet.

\end{abstract}

\section{Introduction}\label{sec:intro}

Spiral galaxies are thought to form inside-out: gas that collapses in the center cools in a short timescale and stars form at an accelerated pace with respect to the outer regions \citep{larson1976}.
Metals, produced by the stellar nucleosynthesis, are injected into the interstellar medium (ISM) during the lifetime of stars, producing the gas chemical enrichment.
The evolution of a galaxy is then regulated by the possible increase of gas content (either through cold gas accretion or by merging with nearby satellites), by the ability of the gas to form new stars and by the possible enriched gas ejection/re-distribution due to the stellar and/or active galactic nuclei feedback (see \citealt{maiolino2019} for a review).
In addition, galaxies suffer from environmental effects that can shape their properties, such as morphology, gas fraction, star formation rate (SFR).
The metallicity of galaxies stems from all these ingredients and, therefore, is a fundamental tracer to explore galaxy evolution.

The global chemical abundance of the gas in disk galaxies linearly increases with the stellar mass according to the mass-metallicity relation (MZR) reaching an asymptotic value at the highest masses \citep{tremonti2004,kewley2008,sanchez2013,blanc2019}.
The secondary dependence of the gas metallicity on the SFR, i.e., at a given stellar mass high-SFR galaxies have lower chemical abundances \citep{mannucci2010,cresci2019}, brought to light the anti-correlation between the ISM metallicity and the gas fraction \citep{derossi2016,brown2018}, that is interpreted as an effect of dilution due to metal-poor gas accretion \citep{delucia2020}.
Also the galaxy environment plays a role, as cluster galaxies are slightly more metal-rich with respect to field counterparts at fixed stellar mass \citep{maier2019b,franchetto2020}, though such enrichment might be due to a local density effect \citep{cooper2008,ellison2009,peng2014,wu2017}.

A way of understanding the mechanisms of formation and evolution of galaxies is investigating the radial distribution of the ISM metallicity across galaxy disks. The first evidence of negative abundance gradients in a small sample of spiral galaxies goes back to \citet{searle1971}, and was confirmed by many subsequent works
\citep{pagel1981,zaritsky1994,deharveng2000,magrini2007,bresolin2007,berg2013,berg2015}. The decrease of the metallicity as a function of the galactocentric distance is often interpreted as the natural result of the inside-out formation of galaxy disks \citep{prantzos2000}.
In addition, signs of flattening of the metallicity profile are found beyond $R_{25}$, \citep{bresolin2015}. On other hand, positive metallicity gradients are measured in some high-redshift galaxies and in merging systems, as a consequence of metal-poor gas accretion and mixing of the gas \citep{cresci2010,rupke2010,rich2012}.

The advent of spectroscopy based on integral-field unit (IFU) technology, able to explore the whole galaxy disk, has boosted galaxy metallicity studies.
Studies based on CALIFA (Calar Alto Legacy Integral Field Area; \citealt{sanchez2012}) confirm the deviations from a pure linear metallicity profiles, finding  a flattening, or even a drop, in the inner region and/or an outer plateau at large radii in a fraction of galaxies \citep{sanchez2014,sanchezmenguiano2016,sanchezmenguiano2018}.
\citet{belfiore2017} and \citet{mingozzi2020} report that, in MaNGA (Mapping Nearby Galaxies at Apache Point Observatory; \citealt{bundy2015}), metallicity profiles normalized by the effective radius tend to steepen with increasing stellar mass up to $10^{10.5}\,{\rm M_\odot}$, in agreement with a small sample of SAMI (Sydney-AAO Multi-object Integral field; \citealt{croom2012}) galaxies \citep{poetrodjojo2018},
and then slightly flatten at higher masses.
Azimuthal variations of the gas metallicity, driven by the spiral arms mixing and exchanging gas and metals with inter-arm ISM, are observed in some PHANGS (Physics at High Angular resolution in Nearby GalaxieS) and TYPHOON galaxies \citep{ho2017,ho2018,kreckel2019,kreckel2020}.

Metallicity gradients stimulate interest to theoretical works that, on the one hand try to interpret the observations and explore the evolution of the gradient slope with redshift \citep{gibson2013,ma2017,hemler2020}, and on the other hand aim at using the measured profiles to verify their recipes. For example, works based on semi-analytical models have tried to explain the diversity of gas metallicity gradients in galaxies, finding that
mechanisms like radial motion, metal-poor and metal-rich gas accretion, turbulent transport, and outflow are involved in shaping the gas metallicity profiles and adding to the data scatter \citep{fu2013,pezzulli2016,collacchioni2020,sharda2021}. Hydrodynamical simulations have exploited observed metallicity profiles to test the reliability of their simulated galaxies \citep{tissera2019,valentini2019}. 
To constrain the chemical evolution of galaxies it is also important to study the relation between metallicity gradients and galaxy properties.
\citet{lutz2021} report that the metallicity gradients positively correlate with the average stellar mass surface density and anti-correlate with the atomic gas mass.

In the literature, this is no general consensus on the role of environmental conditions in shaping the abundance gradients in disc galaxies. Exploiting MaNGA data, \citet{schaefer2019} show that satellite galaxies with intermediate stellar masses are more metal-rich than centrals, but there is no obvious difference in the metallicity gradients. \citet{lian2019} find that MaNGA low-mass satellites in denser environments tend to have flatter gas metallicity gradients and speculate on accretion of enriched material.
In spite of the recent IFU surveys, the environmental dependence of the metallicity gradients is still an open question.

In this paper we study the radial distribution of ionized gas metallicity and stellar mass in the GASP (GAs Stripping Phenomena in galaxies with MUSE; \citealt{poggianti2017}) sample, dividing galaxies according to their environment. 
GASP is an ESO Large Program carried out with the integral-field spectrograph MUSE, that observed more than 100 galaxies with a spatial resolution of 1~kpc and a sky coverage of, on average, 8 effective radii.
At the same time, we exploit a sample of MaNGA galaxies that allow us to strengthen the analysis on a more robust statistical basis. Our goal is to probe the difference due to the environment and find the processes that shape the metallicity profiles.

The paper is organized as follows: in Sec.~\ref{sec:sample}, we explain the criteria adopted to select the galaxies from the GASP and MaNGA samples, whose data analysis are described in Sec.~\ref{sec:method}, together with the method adopted to derive the gas metallicity and the radial gradients. In Sec.~\ref{sec:results}, we present the main results of this paper, which are more widely discussed in Secs.~\ref{sec:discussion} and \ref{sec:discussion_cluster}. Finally, Sec.~\ref{sec:conclusion} summarizes the conclusions of our study.

Throughout this work, we use a $\Lambda$ cold dark matter cosmology with $\Omega_M=0.3$, $\Omega_\Lambda=0.7$, $H_0=70\,{\rm km\,s^{-1}\,Mpc^{-1}}$, and a \citet{chabrier2003} Initial Mass Function (IMF).


\section{Galaxy samples}\label{sec:sample}

\subsection{GASP sample}
The GASP sample consists of 114 late-type galaxies in the redshift range $0.04<z<0.07$, observed in different environments (field, filaments, groups and clusters).
Observations were taken by the spectrograph MUSE \citep{bacon2010} at the VLT, with wavelength coverage between 4800 and 9300~{\AA} with a spectral resolution between $R=1770$ and $3590$, respectively. The field of view is approximately $1'\times 1'$, with a spatial sampling of $0.2''$/pixel and a natural seeing of $\le1''$ (that corresponds to $\sim1$~kpc for galaxies in the sample). The large FoV provides a mean coverage of the sky of 8~${\rm R_e}$, that allows us to probe the entire ionized gas disk.

GASP targets are selected from B-band images of the WINGS \citep{fasano2006} and OMEGAWINGS \citep{gullieuszik2015} cluster surveys, and the PM2GC catalog \citep{calvi2011}, with the intent of identifying gas anomalies in galaxies. Indeed, GASP collects galaxies undergoing a large variety of different physical processes, such as ram pressure stripping \citep[e.g.,][]{poggianti2017,poggianti2019,gullieuszik2017,gullieuszik2020,bellhouse2017,bellhouse2021,moretti2018a}, interaction with the cosmic web \citep{vulcani2019a}, gas accretion \citep{vulcani2018a,vulcani2018c}, but also undisturbed galaxies are present \citep{vulcani2019b,franchetto2020}. However, focusing on single processes is beyond the scope of this work, as we are going to only explore the differences between galaxies in cluster and field galaxies.

From the initial sample, we exclude passive galaxies, that do not show emission lines in the MUSE spectra \citep{vulcani2020a,vulcani2021}, and galaxies with truncated ionized gas disks  \citep{fritz2017,gullieuszik2020}, whose metallicity gradient can not be estimated. Moreover, we also remove galaxies undergoing a merging event \citep{vulcani2021}, for which the radial metallicity profile might have been affected by the interaction. After this selection, we remain with a sample of 56 galaxies in clusters and 29 galaxies in less dense environments (field, filaments, and small groups) that we simply indicate as field galaxies.

\subsection{MaNGA sample}
The MaNGA survey includes galaxies in the redshift range $0.01<z<0.15$, with a galaxy coverage fixed to 1.5~${\rm R_e}$, for the primary sample (2/3 of the MaNGA galaxies), and 2.5~${\rm R_e}$, for the secondary sample.
Observations were carried out by the 2.5~m SDSS telescope \citep{gunn2006} at the Apache Point Observatory. The FoV is simultaneously observed by 17 hexagonal fiber-bundle IFUs that vary in diameter, from $12''$ to $32''$ \citep{drory2015}. Each fiber has a diameter of $2''$ and feeds light into the two dual-channel BOSS spectrographs, that works in the spectral range between 3600 and 10300~{\AA}, with a resolution of $R\sim1400$ at $4000$~{\AA} and $R\sim2600$ at 9000~{\AA}.
For each target, three dithered observations are combined to produce a datacube with square spaxels $0.5''$ in size. The median PSF of the MaNGA datacubes has a FWHM of $2.5''$ \citep{bundy2015}, corresponding to 0.5 -- 6.5~kpc depending on the redshift.

The data used in this work are taken from the MaNGA-DR15 catalog, which includes the outputs of the MaNGA data reduction pipeline \citep[and subsequent updates]{law2016} for 4688 spatially-resolved galaxies. First, we select galaxies with redshift $z<0.08$ in order to better match the redshift range of the GASP sample.
We take into account only galaxies classified as star-forming according to the classification scheme proposed in \citet{belfiore2016}, based on the diagnostic BPT-diagram {[{\sc O\,iii}]$\lambda$5007/H$\beta$} versus {[{\sc S\,ii}]$\lambda\lambda$6716,31/H$\alpha$} \citep{baldwin1981,kauffmann2003,kewley2001}.
In addition, we exclude highly inclined systems (minor-major axis ratio $b/a<0.4$).
The selection leads to a sample of 1789 galaxies.
We further classify the galaxies in sub-samples based on the mass of the host halo reported in the catalog of groups and clusters of SDSS galaxies by \cite{tempel2014}. In order to match the host halo mass distribution of GASP galaxies, we identify 103 MaNGA galaxies in clusters with mass $M_{\rm h}>10^{13.5}\,{\rm M_\odot}$ and 1102 MaNGA galaxies in low-density environments ($M_{\rm h}<10^{13}\,{\rm M_\odot}$), while galaxies in the intermediate-mass halos are kept with no classification.

\section{Data Analysis}\label{sec:method}

\subsection{GASP data analysis}\label{sec:method_GASP}

A detailed explanation of methods employed in the analysis of the GASP data is available in \citet{poggianti2017}.
Briefly, MUSE data cubes were reduced with the most updated version of the MUSE pipeline at the moment of observations \citep{bacon2010}\footnote{\url{http://www.eso.org/sci/software/pipelines/muse}}, corrected for extinction due to our own Galaxy assuming the extinction law from \citet{cardelli1989}, and then, average filtered in the spatial direction with a 5~$\times$~5 pixel kernel, corresponding to $1''=0.7-1.3$~kpc, depending on the galaxy redshift.

For each MUSE spaxel, we apply the spectrophotometric code {\sc sinopsis}, which fits the spectrum with different sets of single stellar population models by S.~Charlot \& G.~Bruzual to reproduce its main features and deliver the stellar-only component and the stellar mass \citep{fritz2017}.
Then, spectra are corrected for the underlying stellar absorption, and their emission-line fluxes and corresponding errors are measured using the IDL software {\sc kubeviz} \citep{fossati2016}, which fits emission lines assuming a Gaussian profile. The correction for the extinction by dust internal to the galaxy is applied adopting the extinction curve of \citet{cardelli1989} and assuming an intrinsic Balmer H$\alpha/$H$\beta$ ratio of 2.86, appropriate for an electron density $n_{\rm e}=100\,{\rm cm^{-3}}$ and electron temperature $T_{\rm e}=10^4\,{\rm K}$ \citep{osterbrock1989}.
Only spaxels with a signal-to-noise ratio $(S/N)>3$ for all involved emission lines are considered.

In this work we focus on the values within the galaxy stellar body, whose boundary is defined as the region containing the spaxels whose underlying H$\alpha$ continuum flux is $\sim1\sigma$ above the background level \citep{gullieuszik2020}. The external gas will be studied in a separated paper (A.\ Franchetto, in preparation).

For each galaxy, the inclination ($i$), the position angle ($PA$) of the disk, and the effective radius of the galaxy ($R_{\rm e}$) are estimated performing an isophotal analysis on the I-band images\footnote{The I-band images are derived by integrating the MUSE data cubes on the Cousin I-band response curve function.}, as done in \citet{franchetto2020}.

We use $i$ and $PA$ to deproject the coordinates of each spaxel onto the disk plane of the galaxy and to compute the galactocentric radius fixing the galaxy center to the peak of the underlying H$\alpha$ continuum flux. The correction for the disk inclination is not applied to galaxies with $i<35^\circ$ to avoid biases due to the uncertainties of the derived correction.

The total stellar mass of each galaxy is derived summing up the stellar mass values in all spaxels belonging to the galaxy disk, as in \citet{vulcani2018b}. We measure the stellar mass surface density by dividing the stellar mass in each spaxel by the pixel area expressed in ${\rm kpc^{2}}$ units and corrected for the galaxy inclination.

For the cluster population, \citet{gullieuszik2020} provide the projected clustercentric distances in units of cluster virial radius ($R/R_{200}$) and galaxy line-of-sight velocities with respect to cluster velocity dispersion ($|\Delta v|/\sigma_{\rm cl}$) that we use in Sec.~\ref{sec:discussion_cluster}.

\subsection{MaNGA data analysis}\label{sec:method_MaNGA}

The reduced MaNGA cubes are analyzed by the MaNGA data analysis pipeline v2.2.1 \citep{westfall2019,belfiore2019} that extracts the emission line fluxes, using a Gaussian fitting per line on stellar-continuum subtracted spectra.
Stellar continuum fitting is performed by the software {\sc pPXF} \citep{cappellari2004} that exploits a linear combination of the templates from the MILES stellar library \citep{falcon2011}.
The line fluxes are corrected for the internal dust extinction adopting the \citet{calzetti1997} attenuation curve with $R_{V}=4.05$ and assuming an intrinsic Balmer decrement ${\rm H\alpha/H\beta}=2.86$. 
Following \citet{mingozzi2020}, we exclude from the further analysis the spaxels with $S/N({\rm H\alpha})<15$ and $S/N<1.5$ for the other involved lines.

For each galaxy, the integrated stellar mass, the elliptical Petrosian effective radius $R_{\rm e}$\footnote{The elliptical Petrosian effective radius is the most robust measure of the photometric properties of MaNGA galaxies provided by the NSA catalog. Therefore, we use this quantity as an alternative to the effective radius adopted for GASP galaxies.}, and the semi-axis ratio are taken from the NASA-Sloan catalog (NSA v1\_0\_1, \citealt{blanton2011}). The deprojected distance of each spaxel is derived taking into account the galaxy inclination, computed from the semi-axis ratio and assuming an intrinsic oblateness of 0.13 \citep{giovanelli1994}.

For 1310 galaxies, we can also derive the integrated gas fraction ($f_{\rm gas}=\log(M_\text{\sc Hi}/M_\odot)$), exploiting the atomic gas masses tabulated in the DR2 catalog of the {\sc Hi}-MaNGA survey \citep{masters2019,stark2021}.

We also take the orbital parameters ($r/R_{200}$ and $|\Delta v|/\sigma_{\rm cl}$) of the cluster galaxies from the \citet{tempel2014} catalog.

\subsection{Gas-phase metallicity radial gradients}\label{sec:method_OHgradients}
In this section, we describe the method used to derive the metallicity of the ionized gas, here defined as oxygen abundance\footnote{We use the oxygen abundance as proxy of the metallicity of the gas (mass fraction of all metals), as it is the most abundant heavy element in the ISM.} $12+\log\,({\rm O/H})$, its radial profile for each galaxy, and its gradient.
We make the reader aware that for the GASP sample, that counts only 85 galaxies and extends to larger galactocentric radii, we carry out a more detailed analysis of the profiles, while for the 1789 MANGA galaxies we perform a simpler method, showing that the two methods are consistent for our purposes as discussed below.
The methods described below are used both for GASP and MANGA, unless otherwise stated. First, we exclude the spaxels dominated by ionizing mechanisms other than star formation, using the diagnostic BPT-diagram \citep{baldwin1981} [{\sc O\,iii}]$\lambda$5007/H$\beta$ versus [{\sc N\,ii}]$\lambda$6583/H$\alpha$ and the separation line of \citet{kauffmann2003}.
Then, the gas-phase metallicity is estimated exploiting the strong emission lines of the ionized-phase of the ISM. We adopt a modified version of the {\sc pyqz} code (\citealt{dopita2013}; \citealt{vogt2015}; F.\ Vogt 2017, private communication) that interpolates the observed flux ratios on the line-ratio model grid [{\sc O\,iii}]$\lambda$5007/[{\sc S\,ii}]$\lambda\lambda$6716,6731 versus [{\sc N\,ii}]$\lambda$6383/[{\sc S\,ii}]$\lambda\lambda$6716,6731 computed by {\sc mappingsIV} \citep{sutherland1993,dopita2013}, and delivers the metallicity and the ionization parameter of the emitting ionized gas. This version of {\sc pyqz} is tested in the range $7.39\le 12+\log({\rm O/H})\le 9.39$ and adopts the solar oxygen abundance of 8.69. 

For GASP galaxies we also derive the error associated to the each metallicity value, as done in \citet{franchetto2020}. This error is calculated as the sum in quadrature of the error given by the propagation of the flux uncertainties, and the systematic error, introduced by the uncertainty of model grids that is usually of $\sim0.1$~dex \citep{kewley2008,dopita2013,blanc2015,mingozzi2020}.

Assuming that galactic disks are axis-symmetric, for each galaxy, we compute the gradient of the azimuthally average metallicity profile, derived by radial binning. The binning allows us to minimize the effects of possible azimuthal variations in the metallicity distribution, and better highlights the radial gradient.
All profiles are scaled by $R_{\rm e}$ of the galaxy to facilitate a fair comparison between galaxies of different sizes. 
Considering the adopted normalizing scale-length, all gradients presented in this work are in units of ${\rm dex\,R_e^{-1}}$, therefore we stress that any consideration on this quantity refers to this chosen notation. 

For GASP galaxies we use radial bins of 0.1~${\rm R_e}$, while for MANGA, that has a lower spatial resolution, we adopt bins of 0.2~${\rm R_e}$. In each radial bin, we compute the mean metallicity and its standard deviation. For GASP galaxies, we also derive the error given by the propagation rules applied to the metallicity uncertainties, which we sum in quadrature with the standard deviation.

The metallicity profiles are not always described by a simple linear behavior, as they can also show deviations, like breaks in the slope either at small and large radii \citep{sanchez2014,bresolin2015,sanchezmenguiano2016}. Typically, in the literature the gradients of the metallicity profiles are computed between 0.5 and 2~${\rm R_e}$ in order to avoid biases due to the presence of these features. However, the exact positions of the breaks are variable and fixing the radial range of the gradient computation could partially affect the results, especially when the sample is small.

Since we can probe the entire ionized gas disk of GASP galaxies and we more finely binned their metallicity profiles than MaNGA ones, for GASP galaxies we compute the metallicity gradients with a more sophisticated method, with no assumptions on the presence of the inner and/or outer deviating behaviors, following the procedure performed by \citet{sanchezmenguiano2018}. In detail, we fit the whole profile with three different parameter functions: (a) a `single-linear' profile, (b) a broken linear profile (`double-linear') with a change in slope, (c) doubly-broken linear profile (`triple-linear') with both an inner and an outer break. The mathematical functions are designed as follows:
\begin{itemize}
\item[a)] The `single-linear' profile is given by : \begin{equation}\label{eq:sl}
    12+\log({\rm O/H})(r)=\alpha_{\rm O/H}\,(r-1)+Z_{R{\rm e}}
\end{equation}
where the radial coordinate $r$ is expressed in units of ${\rm R_ e}$, $\alpha_{\rm O/H}$ is the slope of the metallicity profile;
\item[b)] the `double-linear' profile consists in a broken linear function split at $R_{\rm b}$; at $r<R_{\rm b}$ the function is \begin{equation}
    12+\log({\rm O/H})(r)=\alpha_{\rm in}\,(r-1)+Z_{R{\rm e}},
\end{equation}
while at $r>R_{\rm b}$ it is \begin{equation}
    12+\log({\rm O/H})(r)=\alpha_{\rm out}\,(r-R_{\rm b})+\alpha_{\rm in}\,(R_{\rm b}-1)+Z_{R{\rm e}}.
\end{equation}
If the deviation from the main gradient is located at $r>R_{\rm b}$ , we define $\alpha_{\rm in}\equiv\alpha_{\rm O/H}$; if, instead, it occurs in the inner part, $\alpha_{\rm out}\equiv\alpha_{\rm O/H}$;
\item[c)] the `triple-linear' profile is a doubly-broken linear function with both an inner flattening at $R_{\rm in}$ and an outer plateau beyond $R_{\rm out}$; this function is characterized by three segments, where the middle one, at $R_{\rm in}<r<R_{\rm out}$, is given by Eq.~\ref{eq:sl}, while the segment at $r<R_{\rm in}$ is\begin{equation}
    12+\log({\rm O/H})(r)=\alpha_{\rm in}\,(r-R_{\rm in})+\alpha_{\rm O/H}\,(R_{\rm in}-1)+Z_{R{\rm e}}
\end{equation}
and the part at $r>R_{\rm out}$ is \begin{equation}
    12+\log({\rm O/H})(r)=\alpha_{\rm out}\,(r-R_{\rm out})+\alpha_{\rm O/H}\,(R_{\rm out}-1)+Z_{R{\rm e}}.
\end{equation}
\end{itemize}
The slopes ($\alpha_{\rm O/H}$, $\alpha_{\rm in}$, and $\alpha_{\rm out}$) are in units of ${\rm dex\,R_e^{-1}}$, all radii are expressed in units of ${\rm R_e}$, and $Z_{R{\rm e}}\equiv 12+\log({\rm O/H})(R_{\rm e})$.

A posteriori, we choose only the functional shape that best describes the observed metallicity profile, by applying the corrected Akaike Information Criterion\footnote{With respect to the Akaike Information Criterion \citep[$AIC$;][]{akaike1974}, the $AICc$ introduces a second-order penalty term that better perform in small samples.} \citep[$AICc$;][]{hurvich1989,akaike1974} to the three best fits. The $AICc$ evaluates the goodness of each fit penalizing, at the same time, the loss of degrees of freedom. The lowest $AICc$ value indicates the best function.

The fitting procedure is carried out adopting the python module {\sc emcee} \citep{foreman2013} that performs a Markov Chain Monte Carlo (MCMC) sampling of the multi-dimensional parameter space. The best-fit parameters are derived maximizing the Gaussian likelihood function that takes into account the uncertainties of the data points, and imposing uniform priors in the following ranges: $\alpha_{\rm O/H}$, $\alpha_{\rm in}$, and $\alpha_{\rm out}\in[-1.0,0.5]$, $Z_{R{\rm e}}\in[7.6,9.3]$, $R_{\rm b}$, $R_{\rm in}$, and $R_{\rm out}\in[\min(r),\max(r)]$, where $\min(r)$ and $\max(r)$ refer to the extremities of the metallicity profiles. In addition, we impose that $R_{\rm in}<R_{\rm out}$. From the marginalized distributions of each sampled parameter, we compute the median and the semi-range between the 16th and 84th percentiles.
In Tab.~\ref{tab:tabella}, in Appendix~\ref{app:gradients}, we report the best-fit $\alpha_{\rm O/H}$ values and the corresponding uncertainties for each galaxy in our GASP sample. In this work, we do not use the other best-fit parameters ($R_{\rm b}$, $R_{\rm in}$, $R_{\rm out}$, $\alpha_{\rm in}$, and $\alpha_{\rm out}$), but they are tabulated in Tab.~\ref{tab:tabella2} in Appendix~\ref{app:bestparam}.

\begin{figure*}[th]
\centering
\includegraphics[width=1.5\columnwidth]{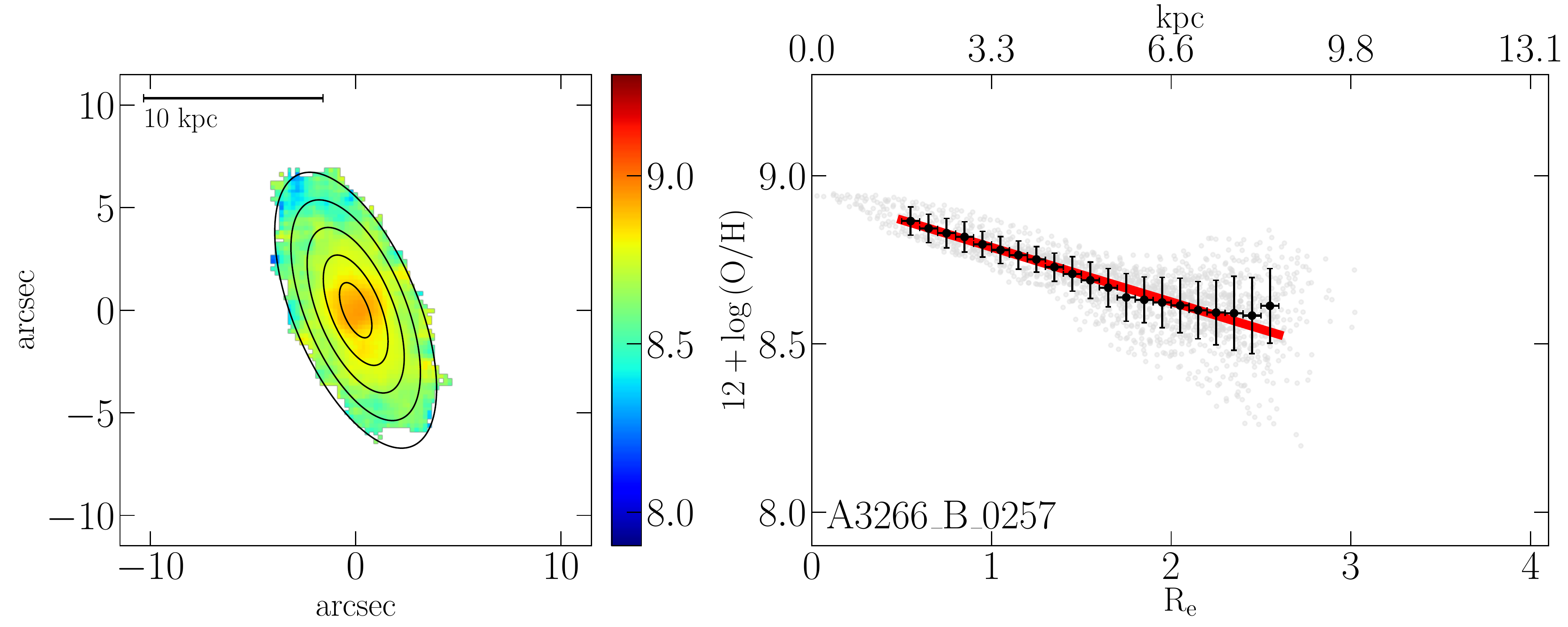}
\includegraphics[width=1.5\columnwidth]{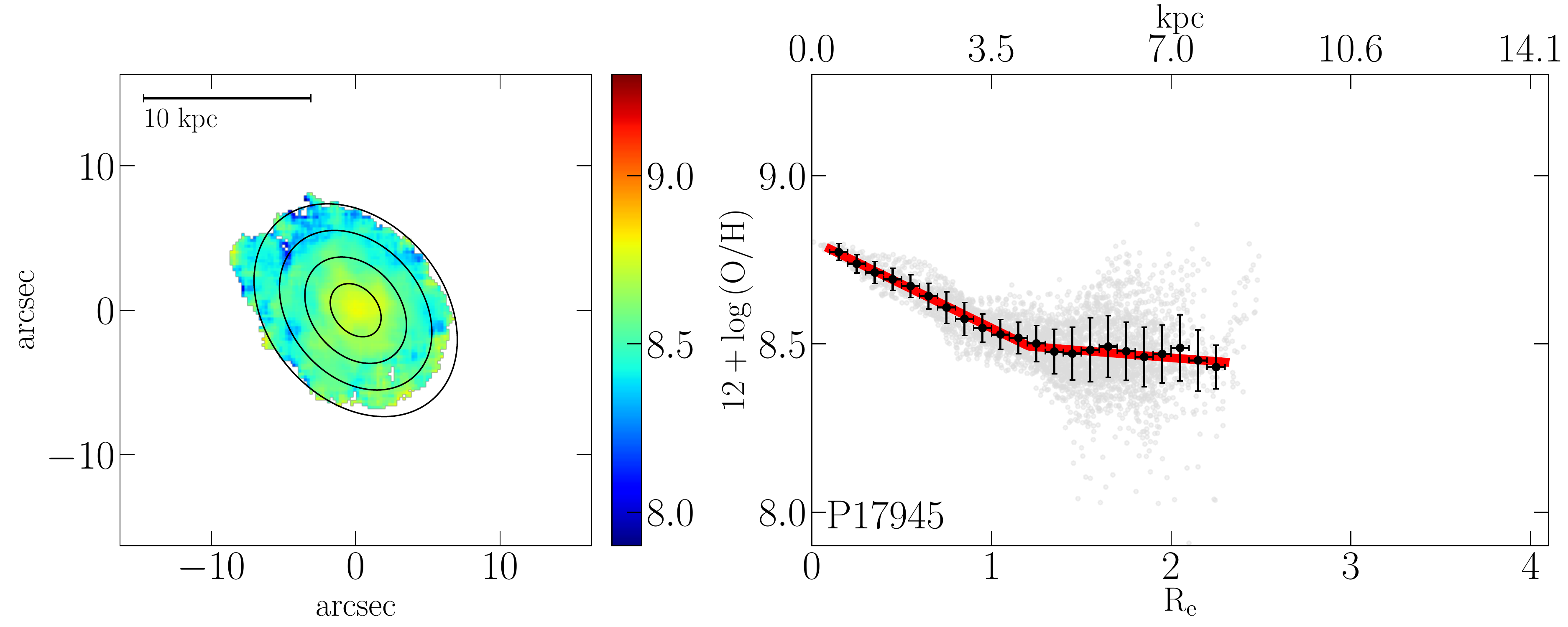}
\includegraphics[width=1.5\columnwidth]{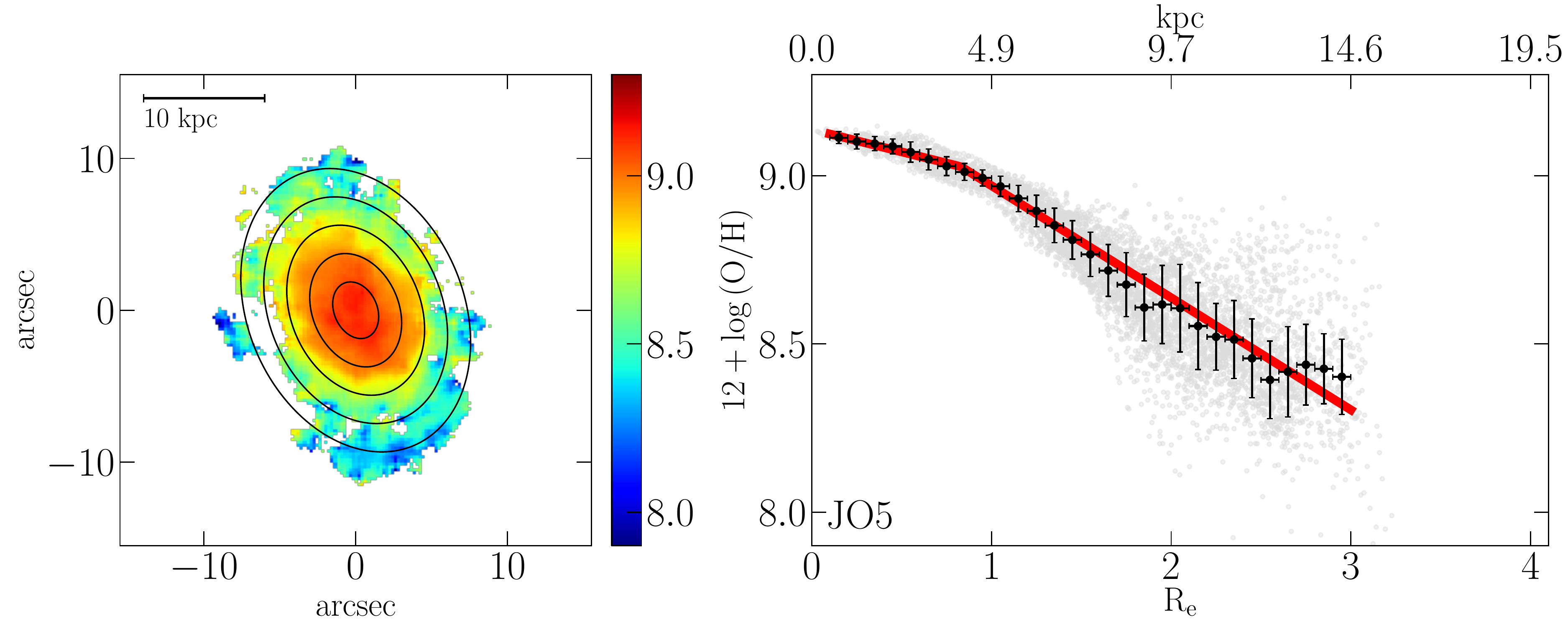}
\includegraphics[width=1.5\columnwidth]{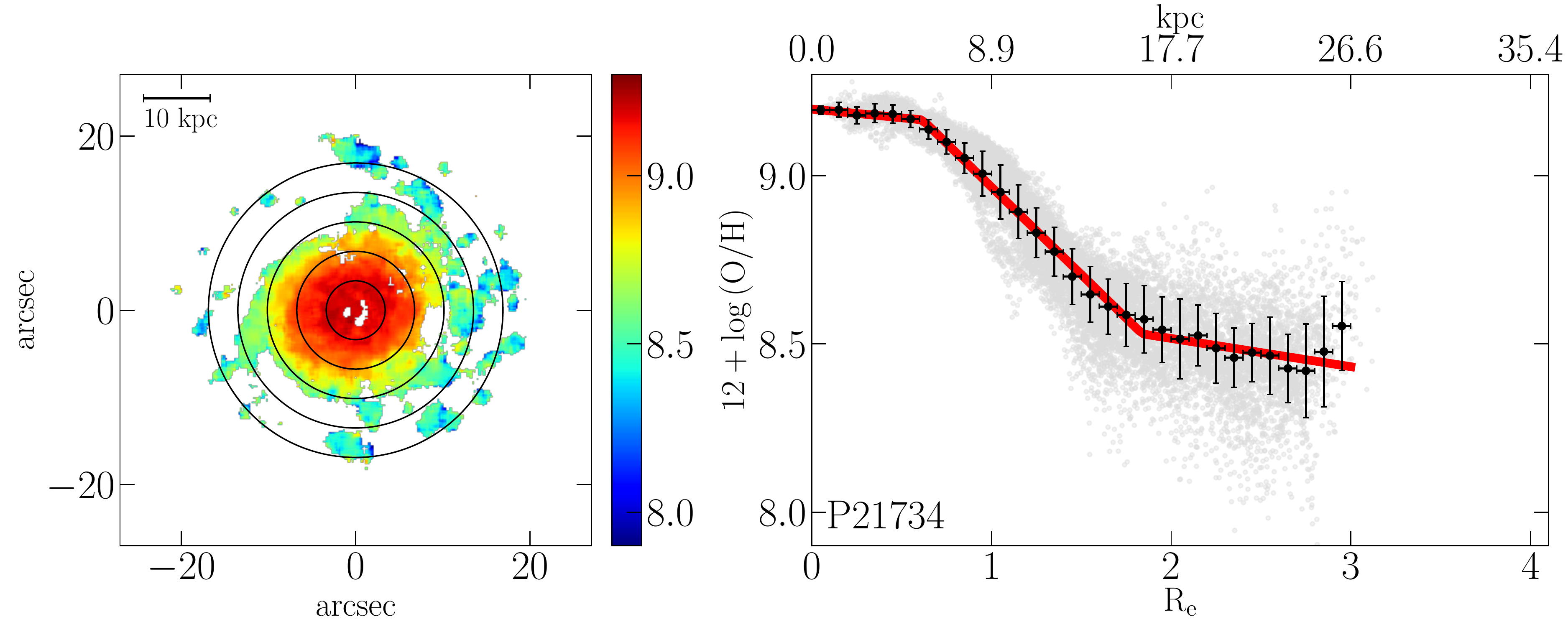}
\caption{Left panels: gas metallicity maps for the galaxy A3266\_B\_0257, P17945, JO5, and P21734. Concentric ellipses indicate the radial distance from the galaxy center, with incremental steps of $0.5\,{\rm R_e}$. We just remark that the PSF $\le 1''$. Right panels: gas metallicity profiles of the  respective galaxies. Gray small points correspond to the spaxel gas metallicities. Black dots show the azimuthally averaged radial profile of the metallicity and the associated uncertainties. Horizontal error bars indicate the size of the radial bins. Red curves are the best fits. The complete figure set of 85 images (one for each galaxy) is available in the online journal. The azimuthally average gas metallicity profiles of each GASP galaxy are available as the Data behind the Figure.\label{fig:OHprofiles}}
\end{figure*}

Figure~\ref{fig:OHprofiles} shows the metallicity maps for four GASP galaxies and the corresponding radial profiles, in order to provide illustrative examples of the variety of metallicity profiles and their best fits. The metallicity profile of A3266\_B\_0257 is represented by a `single-linear' function. P17945 and JO5 show `double-linear' profiles, where the former is characterized by an outer plateau and the latter by an inner flattening. The metallicity profile of P21734 is a clear example of `triple-linear' profile.

Note that the ISM contains a large amount of diffuse ionized gas (DIG, e.g. \citealt{haffner2009}). This warm component is more extended than star forming associations and its ionizing source is still unclear. In some cases, discrepancies between metallicities derived from spectra of {\sc H\,ii} regions and spectra dominated by DIG are reported \citep{zhang2017,poetrodjojo2019,valeasari2019}. \citet{tomicic2021} estimated the fraction of DIG for the galaxies in GASP, exploiting the H$\alpha$ surface brightness and the [{\sc S\,ii}]/H$\alpha$ line ratio measured in each spaxel, and found that the DIG emission contributes between the 20\% and 90\% of the flux. In order to test the robustness of our metallicity gradients against the DIG contamination, we checked that the azimuthally averaged radial profiles of the metallicity do not significantly change if we repeat the computation penalizing the spaxels dominated by DIG emission (the mean variation of the metallicity gradients is within $0.02~{\rm dex\,R_e^{-1}}$). In addition, N.~Tomi\v{c}i\'c et al.\ (in press) statistically look for a difference in metallicity between dense gas and DIG within same radial bins, and found that the difference is lower than the scatter within bins and uncertainties in metallicities.

\begin{figure}[th]
\centering
\includegraphics[width=\columnwidth]{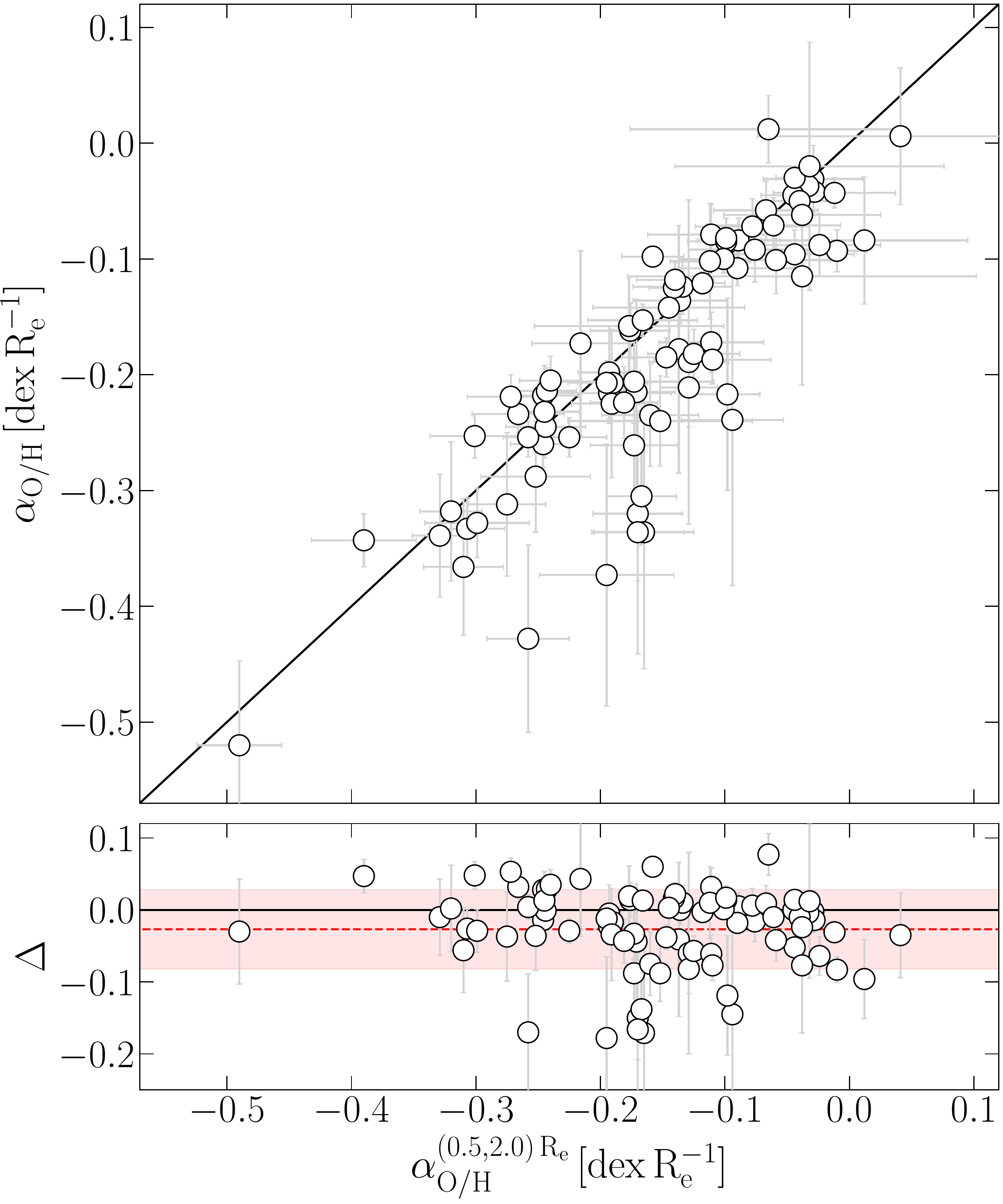}
\caption{Comparison between the metallicity gradients estimated fitting the whole profile with a broken linear function and those derived by the linear fit between 0.5 and 2.0~$\rm R_e$ for the GASP sample. The black line indicates the 1:1 relation. Lower panel: distribution of differences between the two quantities. Red dashed line and faded area show the average offset and the standard deviation, respectively. \label{fig:OHslopes_compare}}
\end{figure}

The method used to calculate the metallicity gradients for GASP galaxies is not suitable for the galaxies in MaNGA sample, as MaNGA observations have a smaller FoV and lower spatial resolution.
Therefore, we fit the metallicity profiles of MaNGA galaxies using Eq.~\ref{eq:sl} in the radial range $0.5-2~{\rm R_e}$. 
To understand how much the choice of the parameterization affects the results, for GASP galaxies we also compute gradients using the same approach adopted for MaNGA. In Fig.~\ref{fig:OHslopes_compare}, we compare the gradients obtained by this method with those given by the previous one. There is a clear correlation, though fitting the metallicity profiles with a linear function in a fixed radial range provides, on average, slightly shallower gradients by $0.027\pm0.055~{\rm dex\,R_e^{-1}}$.

\subsection{Stellar surface mass density gradients}\label{sec:method_Mgradient}
Metallicity well correlates with the stellar mass (e.g.\ MZR, \citealt{tremonti2004}; \citealt{franchetto2020}), and in our analysis we aim at investigating if this connection persists also between the radial gradients of these two quantities. 

For GASP, we compute $\Sigma_\star$ gradients from the stellar mass surface density $\Sigma_\star$ from {\sc sinopsis}.
In principle, we should use  the integral of the star formation rate over the whole galaxy history (the stellar mass ever formed) that should be more closely related to the metals produced. Nonetheless, this quantity requires  a detailed star formation history reconstruction that is not available for MaNGA (see below), so we just use the current stellar mass density instead. We have checked that the two quantities give consistent results.

For MaNGA, we  exploit the maps of stellar mass surface density provided in the MaNGA Pipe3D Value Added Catalog of the SDSS-DR15 \citep{sanchez2016}. The stellar population properties of MaNGA galaxies are derived using the fitting code {\sc fit3d} \citep{sanchez2006} and the single stellar populations template library of \citet{cidfernandes2013} \citep{sanchez2016}. We use these $\Sigma_\star$ maps with a correction factor of 0.24 dex to scale the values down from a Salpeter to a Chabrier IMF.

We derive the azimuthally average radial profiles of $\Sigma_\star$ (expressed in logarithmic scale of ${\rm M_\odot\,kpc^{-2}}$ units), applying the same radial binning adopted for the metallicity (Sec.~\ref{sec:method_OHgradients}).
$\Sigma_\star$ profiles are typically linearly decreasing up to, on average, $2.5\,{\rm R_e}$, with a possible outer plateau (an example of a $\Sigma_\star$ profile is shown in Appendix~\ref{app:Mprofile}). Therefore, these profiles do not manifest the intrinsic complexity observed in the metallicity profiles, and we estimate the gradient of the stellar mass surface density ($\alpha_{M\star}$) by fitting a linear function between 0.5 and 2.0~${\rm R_e}$. This also allows us to avoid possible biases due to the presence of massive bulges, which usually have steeper mass profiles than the disks \citep{gonzalez2014}. We apply a MCMC fitting using the python module {\sc emcee} and imposing the following uniform priors on the free parameters: $\alpha_{M\star}\in[-1.5,1.0]$ and $\Sigma_{\star R{\rm e}}\in[6,10]$, where $\Sigma_{\star R{\rm e}}$ is the stellar mass surface density at $R_{\rm e}$. All $\alpha_{M\star}$ values of GASP galaxies are reported in Tab.~\ref{tab:tabella}.


\section{Results}\label{sec:results}
\subsection{The distribution of the metallicity gradients}\label{sec:OHresults}

\begin{figure*}[th]
\centering
\includegraphics[width=0.47\textwidth]{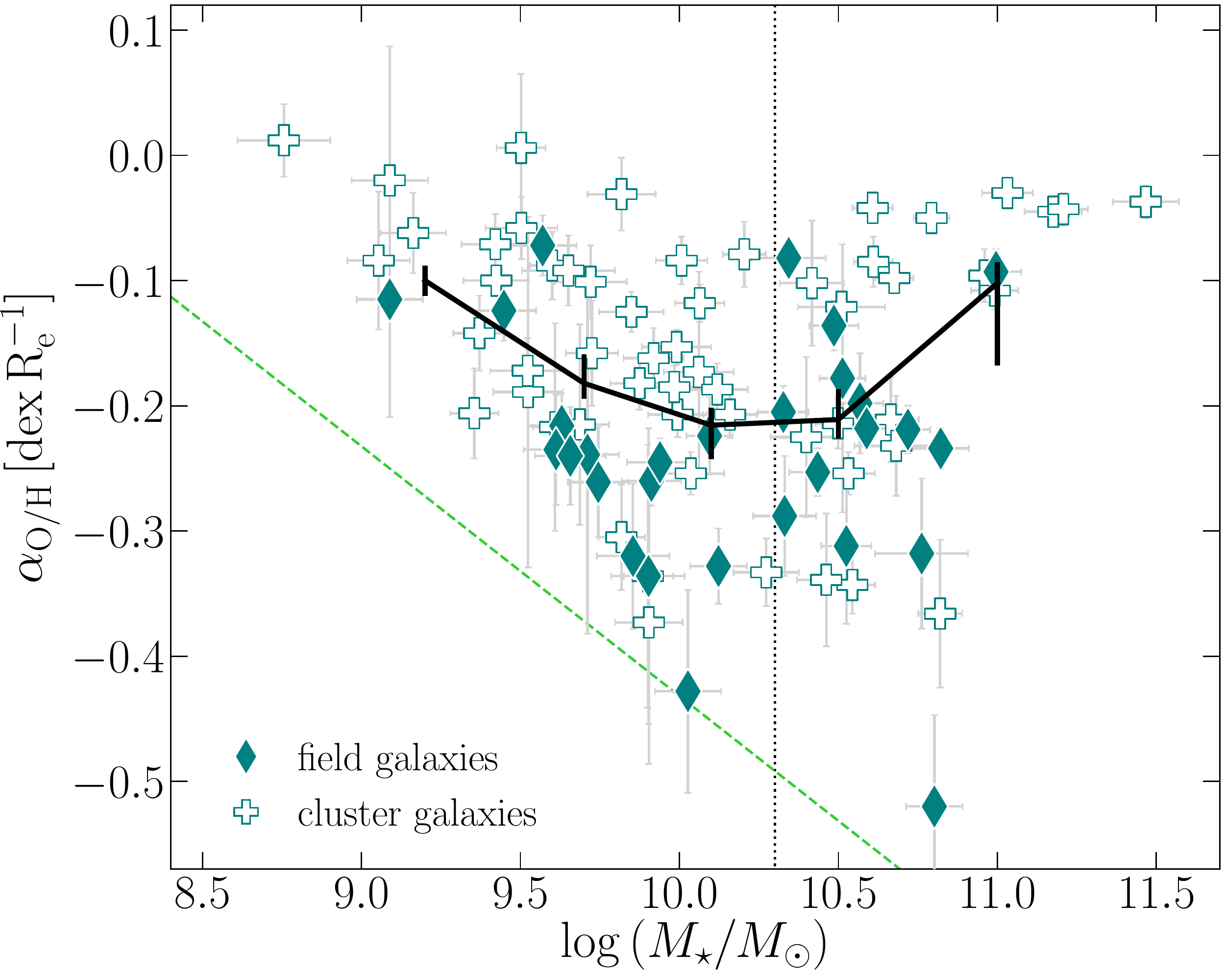}\hspace{0.05\textwidth}
\includegraphics[width=0.47\textwidth]{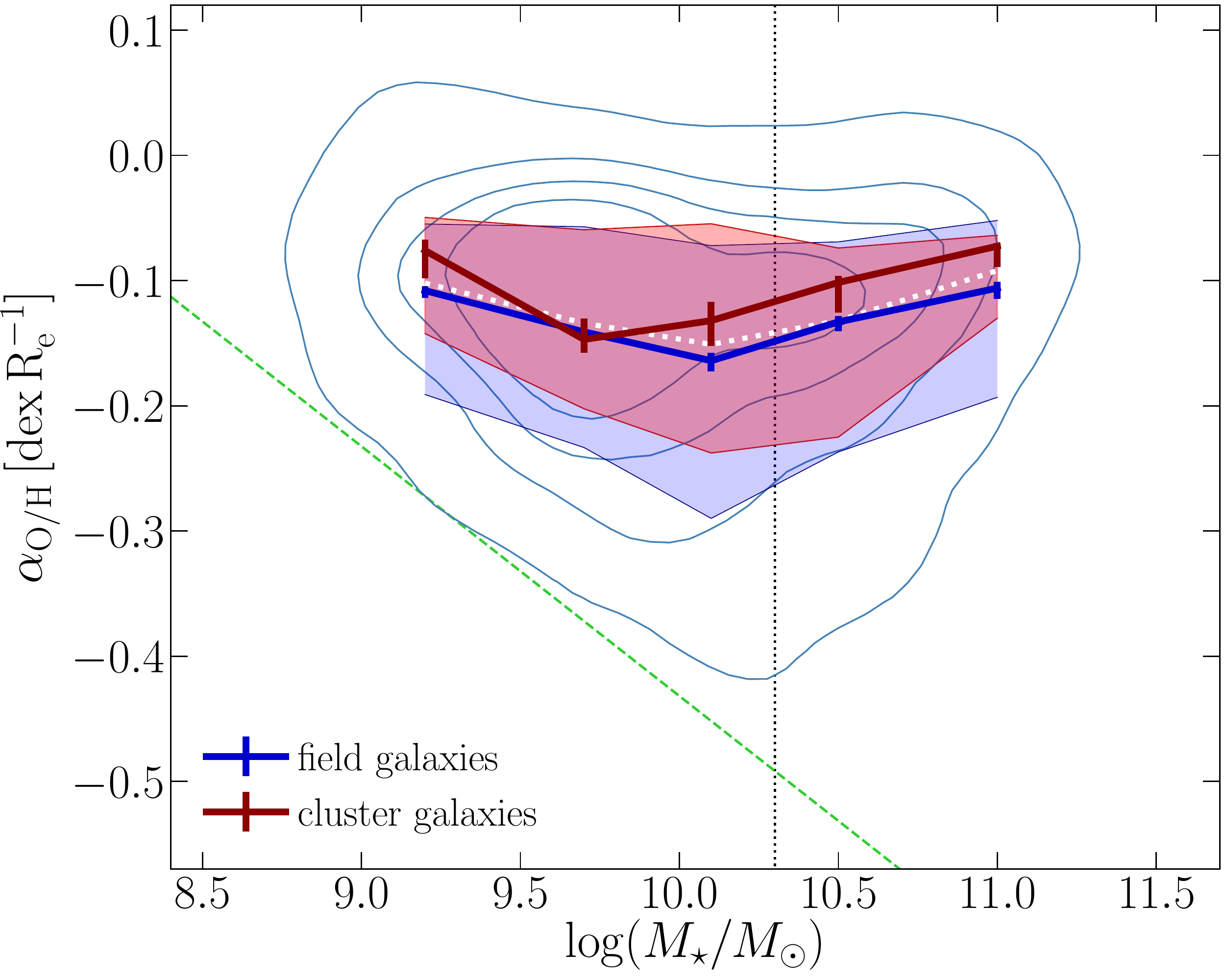}
\caption{Gas metallicity gradient as a function of galaxy stellar mass. Left: distribution of 29 field and 56 cluster GASP galaxies, depicted with filled elongated diamonds and empty crosses, respectively. Black line indicates the median in five stellar mass bins (see the text).
Right: contours enclosing the distribution of the 35\%, 55\%, 75\%, and 95\% of MaNGA galaxies. Dotted white, solid blue, and solid red curves refer to the medians of the whole MaNGA sample, only field galaxies, and only cluster galaxies, respectively, in five stellar mass bins (see the text). The blue and red faded area indicate the range between the 16th and 84th percentiles.
The vertical error bars are the 16th and 84th percentile range divided by $\sqrt{N}$, where $N$ the number of galaxies in the stellar mass bin. In both panels, the vertical dotted line indicates the stellar mass at $10^{10.3}\,{\rm M_\odot}$ and the dashed green line represents the limit of the forbidden region (Eq.~\ref{eq:fb}).
\label{fig:OHslope}}
\end{figure*}

One of the goals of this paper is to investigate the dependence of the gas metallicity gradients on the galaxy environment. Figure~\ref{fig:OHslope} illustrates the distribution of the metallicity gradients of the GASP (on the left) and MaNGA (on the right) samples as a function of stellar mass. We note
that our intent is to explore the same trends in both samples and all the comparisons will be exclusively qualitative.

First, we focus on the left panel corresponding to the GASP sample. 
Before  explaining in detail the results, we list the main observed features: \begin{itemize}
    \item[i)] all metallicity gradients are negative, except for two low-mass galaxies that are positive but close to zero;
    \item[ii)] no low-mass galaxy exhibits a very steep metallicity gradient, but gradients tend to become steeper with increasing stellar mass up to $M_\star\sim 10^{10.3}\,{\rm M_\odot}$;
    \item[iii)] the scatter of the metallicity gradient-mass relation increases for the intermediate masses;
    \item[iv)] the massive galaxies ($M_\star>10^{11}\,{\rm M_\odot}$) show very flat metallicity profiles ($\alpha_{\rm O/H}> -0.05\,{\rm dex\,R_e^{-1}}$);
    \item[v)] at $M_\star<10^{10.3}\,{\rm M_\odot}$, cluster galaxies have, on average, shallower metallicity profiles than field galaxies of similar mass.
\end{itemize}

Negative metallicity gradients trivially mean that the ISM in the central regions of the galaxies has a higher metallicity than the outer gas, as expected according to the inside-out formation of disk galaxies \citep{searle1971}.
We note that the bottom left corner of the diagram shown in Fig.~\ref{fig:OHslope}, corresponding to the region at low masses and lower metallicity gradients, contains no galaxies. This is a sort of \emph{``forbidden region''}, suggesting that low mass galaxies cannot develop very steep metallicity profiles. The indicative limit of this region is given by the equation\begin{equation}\label{eq:fb}
    \alpha_{\rm O/H} = -0.199\,[\log(M_\star/M_\odot)-10]-0.432. 
\end{equation}
This latter was empirically obtained by a linear fitting of the minimum value of metallicity gradients in four stellar mass uniform bins between $10^{9}$ and $10^{11}\,{\rm M_\odot}$, then shifted downward by 0.05~dex to keep all galaxies above the line. This line is not a strict limit, but it just delimits empirically a region where galaxies are not typically found. 

It appears clear that field and cluster galaxies of the GASP sample have different stellar mass distributions.
To minimize possible biases, we divide the GASP sample in two stellar mass ranges, using a separation mass of $\log(M_\star/M_\odot)=10.3$ as a reference. We highlight that a different choice included between $\log(M_\star/M_\odot)=10$ and 10.5 does not significantly impact our conclusions. For galaxies with $M_\star < 10^{10.3}\,{\rm M_\odot}$, we observe an anti-correlation between metallicity gradient and stellar mass (Pearson correlation coefficient $PCC=-0.55;\,\text{p-value}=10^{-5}$). Field galaxies are, on average, 0.09~dex closer to the margin of the forbidden region than cluster galaxies, while cluster galaxies manifest a larger scatter of metallicity gradients spreading above the field population. The Kolmogorov-Smirnov (KS) test confirms with a high probability that the distributions of metallicity gradients of the two populations are different (p-value~$= 0.002$).
For $M_\star>10^{10.3}\,{\rm M_\odot}$, we find a moderate positive correlation with the stellar mass ($PCC=0.38;\,\text{p-value}=0.03$). Both field and cluster galaxies span a wide range of metallicity gradients. At the highest stellar masses, above $\sim 10^{11}\,{\rm M_\odot}$, all galaxies exhibit shallow metallicity profiles ($\alpha_{\rm O/H}>-0.05\,{\rm dex\,R_e^{-1}}$). 

Now, we examine in the right panel of Fig.~\ref{fig:OHslope} the results based on the MaNGA sample. This large sample allows us to increase the statistics and verify the results obtained from the GASP sample. Overall, the distribution of metallicity gradients as a function of the stellar mass is qualitatively in agreement with that of the GASP sample. The main findings are:
\begin{itemize}
    \item[i)] we find an anti-correlation between stellar masses and metallicity gradients up to $10^{10.3}\,{\rm M_\odot}$, and a correlation at higher stellar masses. However, both relations are weaker than those found for the GASP sample;
    \item[ii)] at intermediate masses (around $10^{10.3}\,{\rm M_\odot}$), we observe a large spread (more than 0.3~dex) of metallicity gradient values;  
    \item[iii)] cluster galaxies have, on average, flatter metallicity gradients than field galaxies and, in particular, the population of cluster galaxies has fewer steep metallicity profiles than the population of field galaxies at a given stellar mass.
\end{itemize}

The distribution of the metallicity gradients as a function of stellar mass is similar of that of the GASP sample. At the lowest and the highest masses, galaxies have shallow profiles, while moving toward intermediate masses galaxies can exhibit also steep profiles. No galaxy with steep gas metallicity profile is found at the lowest masses (i.e.\ in the forbidden region), but also at the highest masses, suggesting the presence of a further forbidden region at these masses.

The $PCC$ attests the presence of a mild anti-correlation at $M_\star<10^{10.3}\,{\rm M_\odot}$ ($PCC=-0.27;\,\text{p-value}=10^{-18}$), and a weak correlation at $M_\star>10^{10.3}\,{\rm M_\odot}$ ($PCC=0.25;\,\text{p-value}=10^{-10}$). The weakness of these trends with the mass resides in the large scatter of gradient values which fill all the area above the forbidden regions. In particular, the largest scatter is observed around $M_\star=10^{10.3}\,{\rm M_\odot}$, where 90\% of galaxies with this stellar mass spans more than 0.3~dex. A small fraction of galaxies (3\%) shows inverted profiles (i.e.\ positive gradients), that could be ascribed to a contamination of merging systems \citep{rich2012,grossi2020}.

To study the dependence of the metallicity gradients on the stellar mass, we divide the MaNGA sample in five stellar mass bins delimited by the values $\log(M_\star/M_\odot)=8.9,9.5,9.9,10.3,10.7,11.3$\footnote{The first and last bins are larger to increase the statistics.}, and estimate the median metallicity gradient in each bin. First of all, we consider all 1789 MaNGA galaxies regardless of their environment, finding again a weak relation between metallicity gradients and stellar masses. On average, the metallicity profiles become slightly steeper up to intermediate masses ($\sim 10^{10.3}\,{\rm M_\odot}$) and then flatten toward higher mass values. To provide a qualitative comparison between the GASP and MaNGA samples, we also estimate the median metallicity gradients of the GASP sample in the same stellar mass bins, finding a similar relation, albeit the values are different.

Coming back to the MaNGA sample, we now contrast the field and the cluster populations to highlight possible differences due to the environment. For four out of five stellar mass bins, median gradients of cluster galaxies are $\sim 0.03$~dex shallower than those of field galaxies, albeit the KS test is not conclusive to verify a statistical difference between the two populations.
We find that, in each stellar mass bin, the vast majority of cluster and field galaxies occupy the same region of the $\alpha_{\rm O/H}-\log(M_\star/M_\odot)$ plane, but field galaxies show a tail of galaxies with low metallicity gradients (lower by $\sim 0.05$) that is absent in clusters. 

\begin{figure}
    \centering
    \includegraphics[width=\columnwidth]{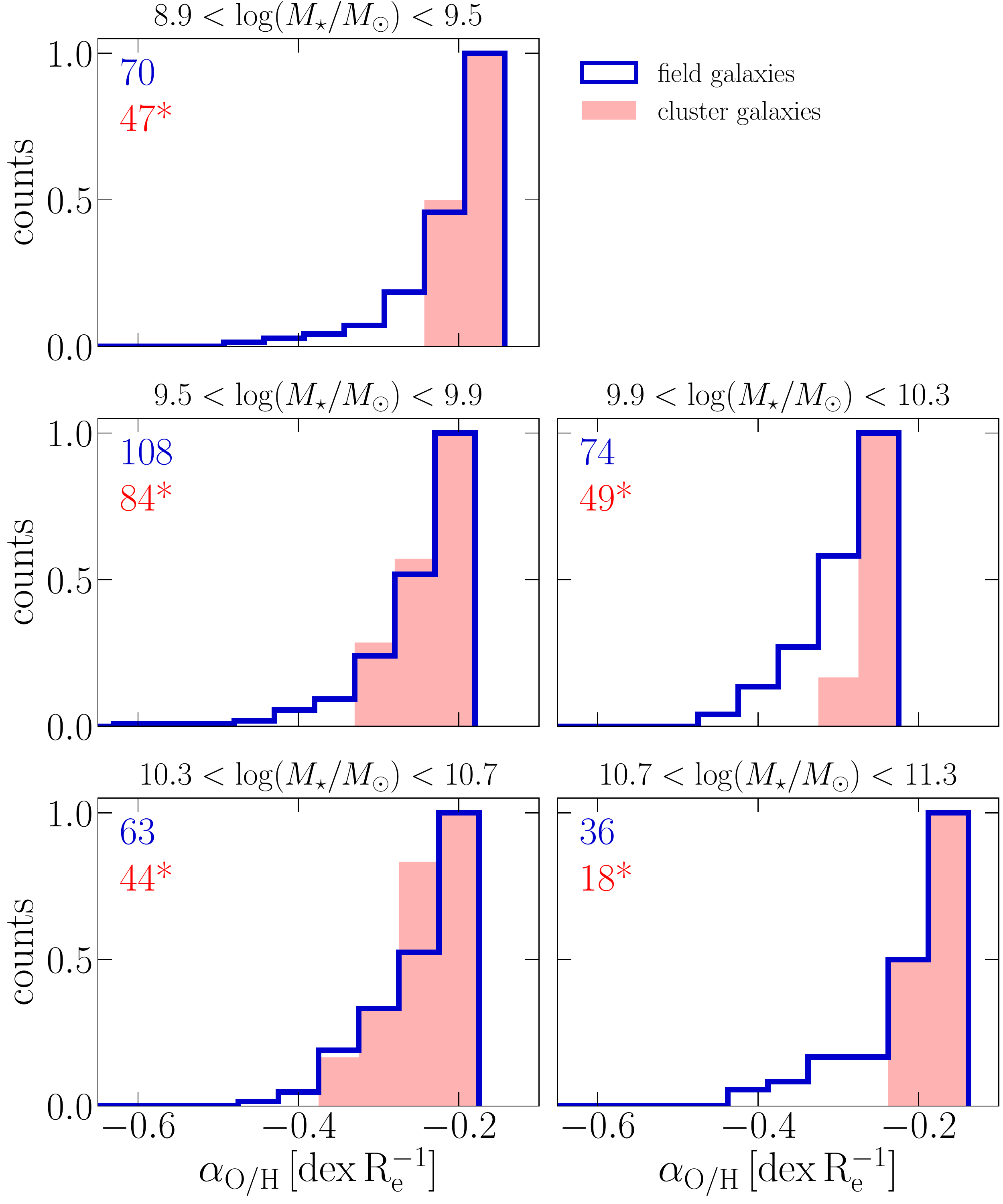}
    \caption{Cumulative distributions in five stellar mass bins of the gradients of the steeper MaNGA metallicity profiles of field (blue) and cluster (red) galaxies than the 33th percentile, in each bin, of metallicity gradients of field galaxies. The numbers in the top-left corner of each panel indicate the number of field (blue) and cluster (red) galaxies within the corresponding stellar mass bin. The asterisk indicates that the number is normalized to take into account that cluster galaxies are fewer than field galaxies by an order of magnitude. \label{fig:manga_steep_aOH}}
    
\end{figure}

In order to examine in detail this difference we select, in each stellar mass bin, those field and cluster galaxies whose metallicity profiles are steeper than the 33th percentile of the metallicity gradients of field galaxies and we show the cumulative distributions of their metallicity gradients in Fig.~\ref{fig:manga_steep_aOH}. In each stellar mass bin, there is a number of field galaxies with steeper metallicity profiles than cluster galaxies. We exclude that this difference is due to either the smaller number of cluster galaxies than the field population or a statistical fluctuation in the number of galaxies. Indeed, if we assume that the number of galaxies is affected by a Poissonian statistical error and taking into account that the cluster galaxies are one order of magnitude fewer than the field population, we find that, in each stellar mass bin, the amount of field galaxies with steep metallicity profiles (i.e.\ metallicity gradients lower than the 33th percentile) is greater than 2$\sigma$ than the number of cluster galaxies. Therefore, we can conclude that cluster members do not show steep metallicity profiles as field galaxies do.

To summarize, GASP and MaNGA galaxies show a characteristic distribution of metallicity gradients as a function of the stellar mass, where low mass and high mass galaxies present shallow metallicity profiles that become steeper moving toward intermediate masses. The cluster population has, on average, flatter profiles than field galaxies at similar stellar masses.

\subsection{The distribution of the stellar surface mass density gradients}\label{sec:Mresults}

In the assumption that the effect of stellar migration \citep{debattista2017,grand2015} and tidal interaction are negligible, the stellar mass gradient provides a proxy for the mass assembly history of a galaxy as it indicates the amount of stars at each radius. Therefore, this quantity can provide a point of view complementary to the metallicity gradients for studying the evolution of galaxies.

\begin{figure}[t]
\centering
\includegraphics[width=\columnwidth]{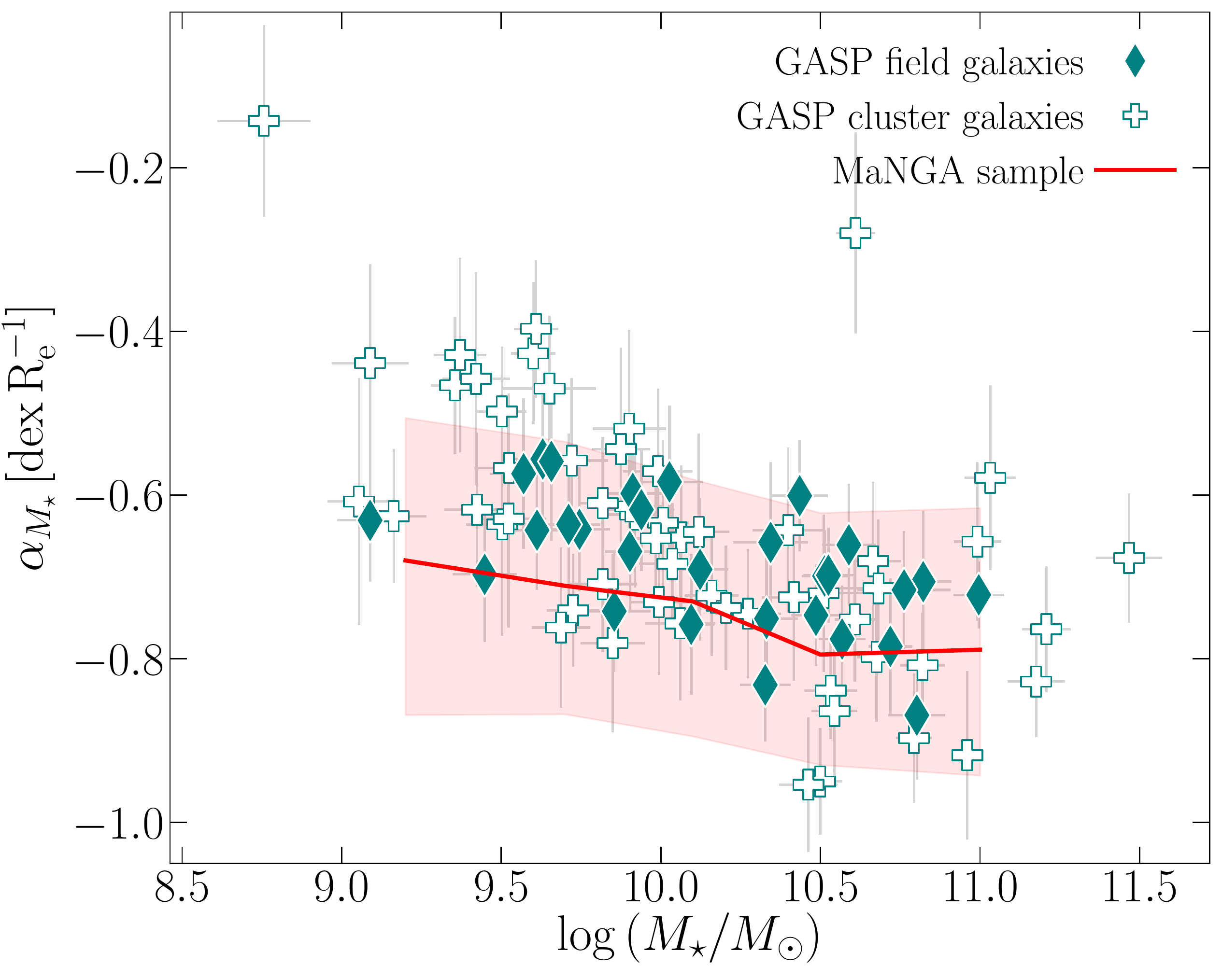}
\caption{Distribution of stellar mass gradients of GASP galaxies as a function of stellar mass. Empty crosses and filled elongated diamonds refer to cluster and field galaxies, respectively. The red line and the faded area indicate the median trend and the range between the 16th and 84th percentiles of MaNGA galaxies. \label{fig:Mslope}}
\end{figure}

Figure~\ref{fig:Mslope} illustrates the stellar mass surface density gradient (stellar mass gradient for brevity) as a function of stellar mass for field and cluster galaxies in the GASP sample.
As expected, all stellar mass gradients are negative, indicating that the central regions are denser than the outer ones. In addition, we observe that: \begin{itemize}
    \item[i)] the stellar mass gradients clearly anti-correlate with the galaxy stellar mass\footnote{Galaxy with shallow stellar mass profile and high stellar mass is JO171: an example of Hoag-type galaxy \citep{moretti2018a}.} ($PCC=-0.59;\,\text{p-value}=10^{-9}$), as already observed in \citet{gonzalez2014};
    \item[ii)] field and clusters galaxies are placed along the same sequence.
\end{itemize}
The lack of offset between field and cluster galaxies suggests that  the radial distribution of the assembly history does not depend on the environment and that the stellar mass profile is probably mainly dominated by stars formed before the galaxy fell in the cluster. In Fig.~\ref{fig:Mslope}  we also show the median trend of MaNGA galaxies along with the 16th and 84th percentiles, computed in the five stellar mass bins adopted in Fig.~\ref{fig:OHslope}. Although the $\Sigma_\star$ profiles of MaNGA galaxies are, on average, slightly steeper than GASP ones, we qualitatively find a similar trend. 
We stress that the $\Sigma_\star$ gradients of the two samples can not be directly compared because the procedures adopted by the two surveys to derive the stellar masses presents substantial differences; for instance, the different sets of single stellar population models used to fit the spectra. Moreover, conversely to GASP data, MaNGA data require a spatial binning to reach a $S/N>50$.


\section{Mass dependence of metallicity gradients}\label{sec:discussion}

In the previous section, we showed that the gas metallicity gradients of both GASP and MaNGA samples are characterized by some remarkable properties: (i) all gradients are consistent with negative values, i.e., the inner regions are more metal-rich than the outskirts; (ii) the metallicity profiles become steeper with increasing stellar mass up to $10^{10.3}\,{\rm M_\odot}$, while at higher masses the metallicity profiles become shallower; (iii) at intermediate stellar masses ($\sim10^{10.3}\,{\rm M_\odot}$), the scatter significantly increases; (iv) stellar mass profiles steepen monotonically with the stellar mass. In this section, we want to understand the physical reasons that are shaping the metallicity profiles: which scenario may explain the steepening of the metallicity profiles as the stellar mass increases up to intermediate masses, also in light of the variation of the stellar mass gradients with galaxy mass? Why do massive galaxies tend to have shallow profiles?

The scenario in which metallicity gradients depend on the stellar mass is qualitatively in agreement with other works in the literature based on MaNGA, SAMI and CALIFA data, adopting different metallicity calibrators \citep{perezmontero2016,belfiore2017,poetrodjojo2018,mingozzi2020}. Some authors, using the CALIFA sample, do not detect any mass dependence and, instead, point to a universal metallicity gradient \citep{sanchez2014,sanchezmenguiano2016}.
However, from the values of metallicity gradients and stellar mass published in \citet{sanchezmenguiano2016}, we find a correlation for $M_\star>10^{10.3}\,{\rm M_\odot}$ ($PCC=0.35;\,\text{p-value}=0.001$), consistently with our results.

\subsection{Star formation as primary regulator of metallicity profiles} \label{sec:discussion_MaOH}

\begin{figure}[t]
\centering
\includegraphics[width=\columnwidth]{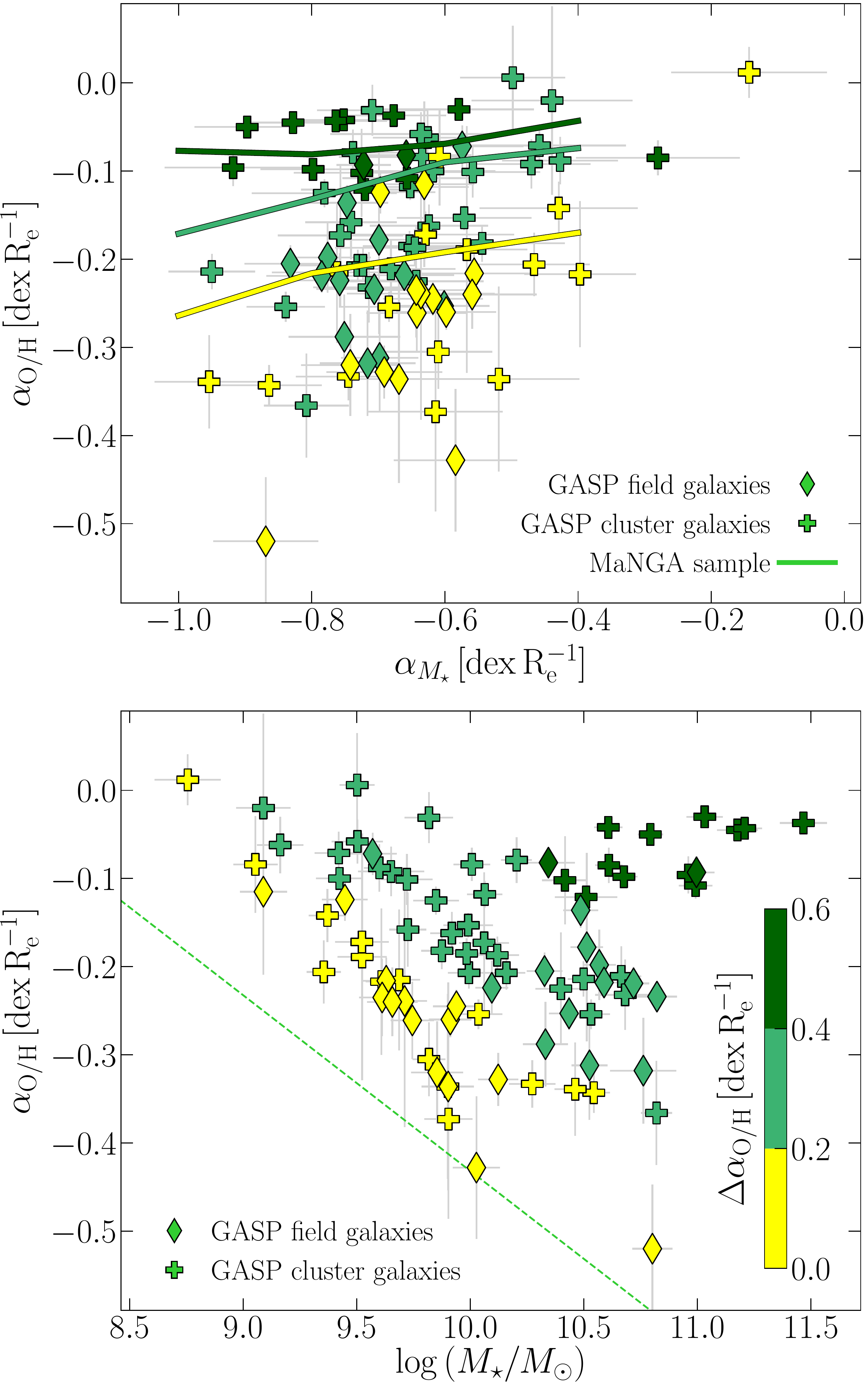}
\caption{Distribution of metallicity gradients as a function of the stellar mass gradient (upper panel) and the stellar mass (lower panel) for GASP galaxies. Elongated diamonds and crosses refer to field and cluster galaxies, respectively, color-coded by the distance $\Delta\alpha_{\rm O/H}$ from the forbidden region (Eq.~\ref{eq:fb} and dashed green line in the lower panel). For a qualitative comparison, the three lines in the upper panel refer to the median trends of MaNGA galaxies, with the same color code used for GASP galaxies.}\label{fig:Mslope_OHslope}
\end{figure}

The inside-out formation implies that the central regions of a disk galaxy rapidly grow, becoming more metal-rich and more massive in stars  per unit area than the peripheral part of the galaxy, and producing radial negative gradients in stellar mass and metallicity \citep{larson1976,prantzos2000}. Keeping this in mind, here we want to investigate the physical reason behind the mass-dependence of the metallicity gradients, and we start considering the in-situ chemical enrichment due to the star formation along the galaxy disk. If we assume that there is no radial motion of gas, the increase in metallicity is directly proportional to the number of stars formed at each radius. The natural expectation is that the metallicity gradients will correlate with stellar mass gradients: the larger amount of stars formed at a given location in the galaxy, the highest the metallicity we should observe. We can test this prediction using GASP galaxies and in the top panel of Fig.~\ref{fig:Mslope_OHslope} we compare the two gradients. Despite the large scatter, we find a mild correlation ($PCC=0.28;\,\text{p-value}=0.009$). In addition, we observe that, while flat metallicity profiles can be found in the presence of both steep and flat stellar mass profiles, the steep metallicity profiles can be found only in galaxies with steep stellar mass profiles. 
In the bottom panel, we propose again the distribution of metallicity gradients as a function of stellar mass of the GASP galaxies. To separate galaxies with similar stellar mass, but different metallicity gradient, for each galaxy we compute the distance $\Delta\alpha_{\rm O/H}$ from the forbidden region (Eq.~\ref{eq:fb} and dashed green line in the figure) and color code the galaxies in both panels of Fig.~\ref{fig:Mslope_OHslope} according to this quantity.
The closest galaxies to the forbidden region are depicted by yellow symbols; the steepening of their metallicity profiles goes at the same pace as the stellar mass profiles steepen, and indeed we find a strong correlation ($PCC=0.60;\,\text{p-value}=0.0005$). On the other hand, the massive galaxies color-coded by dark green, have flat metallicity profiles ($\langle\alpha_{\rm O/H}\rangle=-0.07\,{\rm dex\,R_e^{-1}}$) although their stellar mass profiles are, on average, steep ($\langle\alpha_{M\star}\rangle=-0.71\,{\rm dex\,R_e^{-1}}$). 

In order to verify this behavior for MaNGA galaxies, we probe the distribution of their gas metallicity gradients against the stellar mass gradients, and we show in the top panel of Fig.~\ref{fig:Mslope_OHslope} the median values computed in four equally spaced stellar mass gradients bins dividing the galaxies according to their $\Delta\alpha_{\rm O/H}$ values. Although we restate that the two samples are not quantitatively comparable, overall, we find the same trends observed in the GASP sample, even if the MaNGA sample presents a larger scatter and the correlations are milder, but significant (on average $PCC=0.20;\,\text{p-value}\ll 0.05$).

The emerging picture suggests that the in-situ star formation, and the subsequent metal enrichment,
plays a primary role in steepening the profiles of stellar mass and, consequently, metallicity in disk galaxies. This produces, firstly, the correlation between stellar mass gradients and metallicity gradients and, secondly, the anti-correlation between gradients and stellar masses. This appears to be the dominant mechanism regulating the metallicity profiles up to intermediate stellar masses.
However, the metallicity of the gas in not only determined by the enrichment due to the star formation, but it is also connected to other parameters that might contribute to flatten the metallicity profiles and can be dominant in clusters and at high stellar masses, where we observe the most significant deviations.

\subsection{Dependence on the gas fraction}\label{sec:discussion_deviation}

Overall, our results show a steepening (up to $\sim-0.4\,{\rm dex\,R_e^{-1}}$) of the metallicity profiles with increasing stellar mass up to $\sim 10^{10.3}\,{\rm M_\odot}$. However, many galaxies present shallow ($<-0.1\,{\rm dex\,R_e^{-1}}$) profiles (especially in clusters) at every mass, and an overall flattening of all profiles toward high masses.

\citet{belfiore2017} find that galaxies with lower specific SFR have flatter metallicity profiles and suggest that the metallicity flattening in the central regions of massive galaxies could be due to the general behavior of evolved galaxies to reach an equilibrium abundance at late times, corresponding to low gas fractions. The connection between metallicity and the gas content is well-established: \citet{brown2018} demonstrate that at a given stellar mass, the metallicity decreases as the neutral gas mass increases (see also \citealt{stark2021}).
Moreover, \citet{ascasibar2015} observe that the relation between metallicity and gas fraction also holds at local (1~kpc) scales.
\citet{carton2015} study 23 {\sc H\,i}-rich galaxies, finding that they have steeper metallicity profiles than galaxies with normal {\sc H\,i} gas content and similar mass. \citet{lutz2021} show that metallicity gradients anti-correlate with the atomic gas mass fraction.
\citet{boardman2021} report that at a given stellar mass MaNGA galaxies with steeper metallicity profiles are larger in size, and \citet{sanchezalmeida2018} argue that galaxies  bigger in size have experienced late gas accretion and have lower metallicities than more compact counterparts.
\citet{collacchioni2020}, using cosmological hydrodynamical simulations, conclude that galaxies with high rates of gas accretion have steep metallicity profiles in the inner regions.
Analytic models show that radial motions induced by gas accretion steepen the profiles of chemical abundances \citep{pezzulli2016}.

\begin{figure}[t]
\centering
\includegraphics[width=\columnwidth]{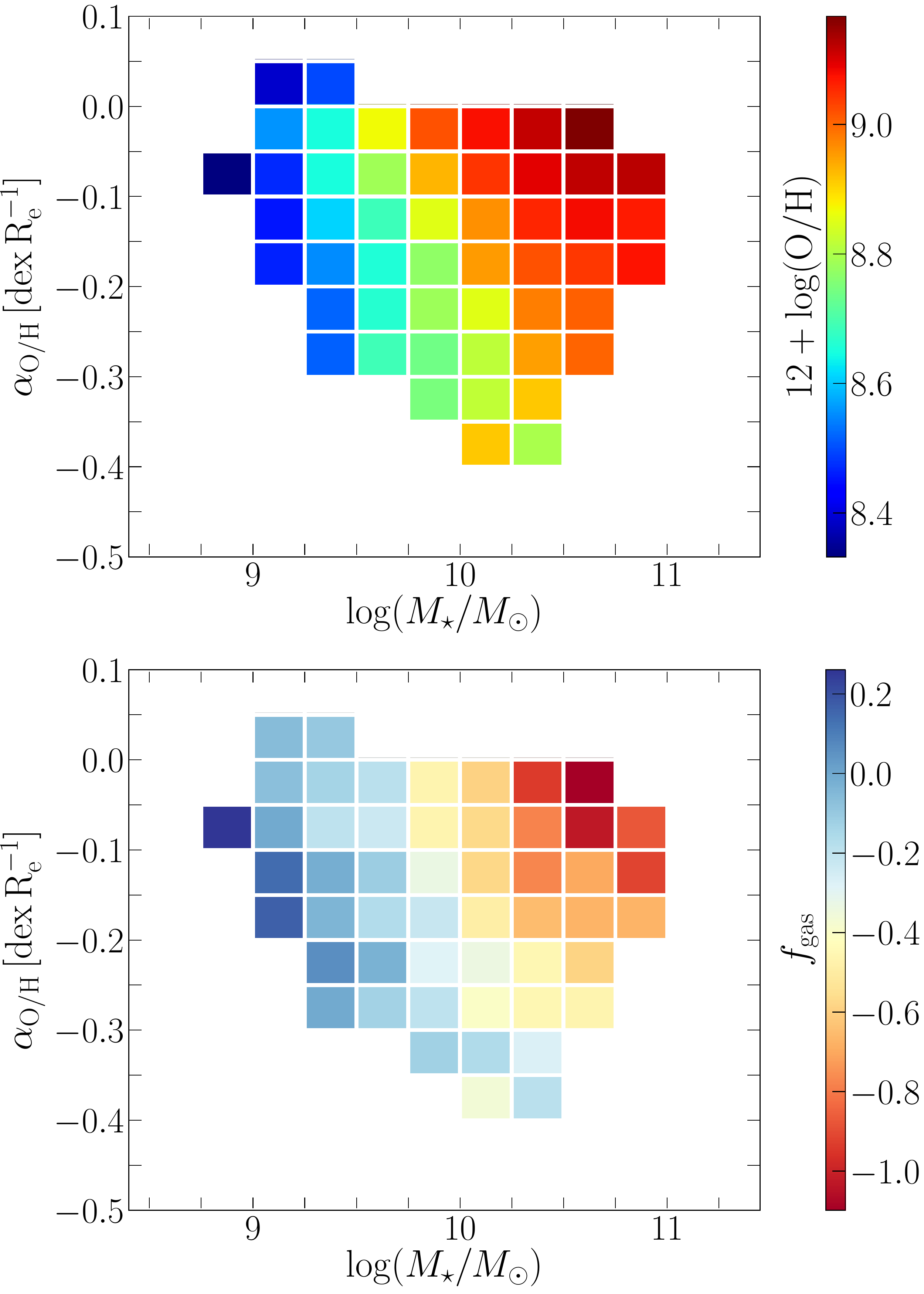}
\caption{Distribution of metallicity gradients of MaNGA sample as a function of the stellar mass, color-coded by the mean metallicity at the effective radius (upper panel) and the mean gas fraction (lower panel). Galaxies are binned in stellar mass (size of 0.25~dex) and metallicity gradient (size of 0.05~${\rm dex\,R_e^{-1}}$). \label{fig:manga_OHslope_fgas}}
\end{figure}

In light of the connection between metallicity and gas content, we investigate the dependence of the metallicity gradients on the gas fraction. To this aim, we exploit the MaNGA sample, for which we have the information on the {\sc H\,i} content.
In Fig.~\ref{fig:manga_OHslope_fgas}, we show the metallicity gradients versus stellar mass, where galaxies are binned in stellar mass (bins of 0.25 dex) and metallicity gradient (bins of 0.05~$\rm dex\,R_e^{-1}$). Each bin contains at least five galaxies, for which we calculate the mean metallicity at $R_{\rm e}$ (upper panel) and the mean gas fraction (lower panel). Moving from left to right, as the stellar mass increases the metallicity increases and the gas fraction decreases. 
In addition, moving from the bottom to the top, at a given stellar mass the metallicity profiles tend to become flatter as the metallicity increases and the gas fraction decreases. 
Galaxies with a high gas content ($f_{\rm gas}>-0.2$) show a tight anti-correlation between stellar mass and metallicity gradient, supporting the steepening of the metallicity profiles with the stellar mass for galaxies whose star formation is in equilibrium with the available amount of gas.

Given the clear presence of a further dependence of the gas metallicity gradients on a third parameter (gas fraction or gas metallicity at $R_{\rm e}$), in addition to the stellar mass, we fitted a three-dimensional surface on the three variables. First, for the two alternative parameters, we bin galaxies in stellar mass (bins of 0.25 dex) and gas fraction (bins of 0.2 dex), and in stellar mass and gas metallicity (bins of 0.1 dex), and we derived the mean gas metallicity gradient in each bin (bins with less than five galaxies are excluded). Then, we fit the mean values with a 2nd-order polynomial surface deriving the relations:
\begin{align}
    \alpha_{\rm O/H}=&-0.46x+0.02x^2+0.83y-0.04y^2\notag\\&-0.09xy+2.61\label{eq:aOH_M_fgas};\\
    \alpha_{\rm O/H}=&-0.76x+0.003x^2-6.10z+0.32z^2\notag\\&+0.07xz+29.89\label{eq:aOH_M_OHre},
\end{align}
where $x=\log(M_\star/M_\odot)$, $y=f_{\rm gas}$, and $z=12+\log({\rm O/H})$ at $R_{\rm e}$. The root mean square of the deviations (RMSD) for Eq.~\ref{eq:aOH_M_fgas} are 0.021 and 0.084 considering the mean metallicity gradients and all MaNGA galaxies, respectively; while, for Eq.~\ref{eq:aOH_M_OHre}, $RMSD=$ 0.034 and 0.078. Both proposed relations are able to follow the shape of the data and completely remove the dependencies of the residuals. The two surfaces are shown in Fig.~\ref{fig:surfaces} in Appendix~\ref{app:surfaces}.

To conclude, our result provides the evidence that, as galaxies consume or lose their gas reservoir, the metallicity increases, in particular at large radii causing the flattening of the metallicity profile.

\subsection{Brief summary}

Thanks to our analysis of the metallicity gradients as a function of the stellar mass gradients and the galaxy gas fraction, we are in the position to propose an explanation for the variety of metallicity gradients found in galaxies of different masses:\begin{itemize}
    \item[i)] Low-mass galaxies: albeit these systems have a large gas fraction, they are not able to produce steep stellar mass profiles, thus their metallicity profiles tend to remain flatter than more massive galaxies;
    \item[ii)] Intermediate-mass galaxies: for these galaxies the radial variation (from the center to the outskirts) of star formation is larger producing steep stellar mass profiles, but they can present a large variety of metallicity gradients. Indeed, as long as the consumption of the gas due to the star formation is balanced by new gas supply, the metallicity profiles can steepen in accordance with the stellar mass gradients; instead, if the gas fraction decreases, the metallicity profiles tend to flatten;
    \item[iii)] High-mass galaxies: although these systems have very steep stellar mass profiles, they are in an advanced stage of their evolution, where most of their gas content was consumed and the metallicities are high at all radii entailing flat metallicity profiles.
\end{itemize}

The balance between in-situ enrichment by stars ever formed at different galactocentric radii and the gas reservoir can give an explanation of the evolution of the gas metallicity profile with the stellar mass in disk galaxies. However, we need to keep in mind that other processes can contribute to shape the metallicity profiles in these systems: accretion of metal-poor gas leads to steeper metallicity profiles because produces a metal dilution effect that amplifies the radial metallicity variation \citep{yates2020}; on other hand, accretion of pre-enriched gas tends to produce flatter metallicity profiles \citep{fu2013,yates2020}; outflows by stellar feedback that mix with the hot halo and fall back into the galaxy disk in the form of galactic fountains with lower specific angular momentum, triggering radial motions, can steepen the metallicity profiles \citep{pezzulli2016}; turbulent radial transport, dominant in low-mass systems, can contribute to maintain flat metallicity profiles \citep{sharda2021}; strong galaxy interactions and major mergers produce almost flat metallicity profiles \citep{rich2012}.

\section{Behaviors of cluster galaxies}\label{sec:discussion_cluster}

In Sec.~\ref{sec:OHresults} we observed that the metallicity gradients of the cluster galaxies are, on average, flatter than those of field galaxies. In this section we attempt to understand this behavior. 
Members of galaxy clusters are usually characterized by higher metallicities and gas deficiency with respect to field galaxies \citep{boselli2006,cooper2008}. These features are interpreted as the result of the suppression of gas accretion, as a consequence of ram pressure stripping \citep{gunn1972}, starvation \citep{larson1980}, or thermal evaporation \citep{cowie1977}. 
The metallicity profiles of cluster galaxies could be interpreted in the context of the previous section where we discussed that galaxies with lower gas fraction show shallower metallicity profiles than galaxies with higher gas fraction and similar mass.
Accretion of enriched material is 
sometimes invoked to explain the behavior of cluster galaxy metallicity profiles \citep{lian2019,schaefer2019}. The accelerated galaxy evolution that occurs in clusters \citep{guglielmo2015} might facilitate this flattening. More rarely, confinement pressure exercised by the intracluster medium could prevent the stellar outflows from leaving the galaxy \citep{mulchaey2010}, making the chemical enrichment of the ISM more efficient.

However, if we take into account that galaxy clusters host various types of galaxy populations, which differ in orbital parameters and time of infall in the halo potential, we cannot simply generalize the behavior of the metallicity gradients observed in cluster galaxies. The GASP sample offers the opportunity to disentangle this issue, as the identification of sub-populations is already provided in many works of the GASP series \citep[e.g.][]{vulcani2019b,vulcani2020b,franchetto2020}.\footnote{Such detailed distinction is instead not available for MaNGA.} In particular, we use the classification adopted in \citet{franchetto2020}, who study the MZR of three sub-samples of GASP galaxies: reference galaxies in clusters (16 objects) and reference galaxies in the field (15 objects) --- whose distribution of H$\alpha$ emission coincides to the stellar morphology and that does not show evidence of gas anomaly features, and cluster galaxies undergoing different stages of ram pressure stripping (RPS; 29 objects), clearly identified by the morphology of their H$\alpha$ emission that presents gas tails and debris or collimated gas extent unilaterally displaced beyond the stellar disk, whose stellar kinematics is undisturbed.
This sub-selection excludes the field and cluster galaxies whose peculiar morphologies  are dealt with in \citet{vulcani2021} and B.~M.~Poggianti (in preparation), respectively.

\begin{figure}[th]
\centering
\includegraphics[width=\columnwidth]{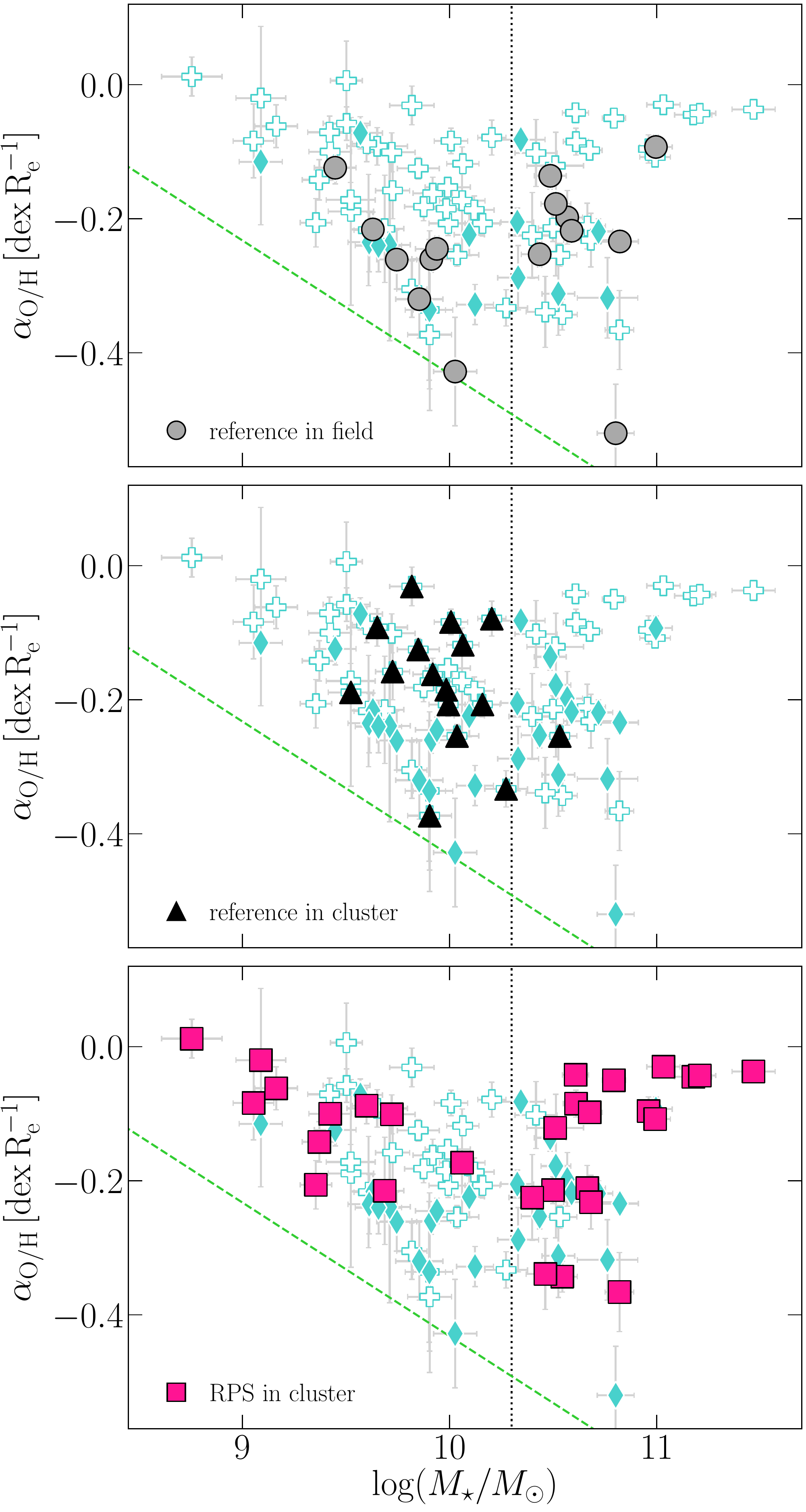}
\caption{Distribution of the metallicity gradients  as a function of the stellar mass. From the top to the bottom panels, we highlight reference field galaxies (gray dots), reference cluster galaxies (black triangles), and stripping cluster galaxies (pink squares), while in each panel, filled elongated diamonds and empty crosses indicate all field and cluster GASP galaxies, respectively, vertical line and dashed green line are as in Fig.~\ref{fig:OHslope}. \label{fig:OHslope_samples}}
\end{figure}

In Fig.~\ref{fig:OHslope_samples}, we explore the distribution of metallicity gradients for these three GASP galaxy categories. 
Since the stellar mass distributions of these samples are different \citep{franchetto2020}, we can not make a robust comparison. However, we can provide general considerations.
In the top panel we focus on the reference field galaxies: their metallicity profiles steepen with the stellar mass up to intermediate masses, while at high masses no clear sequence is detected.
In the middle panel, we highlight the reference cluster galaxies: they are distributed in a smaller stellar mass range ($\rm 10^{9.5}<M_\ast [M_\odot]<10^{10.5}$) than the other samples, and their metallicity profiles are, on average, flatter than those of field galaxies.
In the bottom panel, we consider the galaxies affected by ram-pressure stripping: although these galaxies suffer the gas removal due to the interaction with the intracluster medium, their metallicity profiles seem consistent with the trend observed for other GASP galaxies and the MaNGA sample, and show the steepening toward intermediate stellar masses and a subsequent flattening at high masses. This indicates that the RPS does not strongly affect the metallicity profiles, since the gas removal occurs on the galaxy outskirts and, then, proceeds outside-in, leaving the gas that is not still stripped mostly unchanged. 
We also note that most of the massive RPS galaxies host a central AGN, whose involved processes might influence their metallicity profile.

\begin{figure}[t]
\centering
\includegraphics[width=\columnwidth]{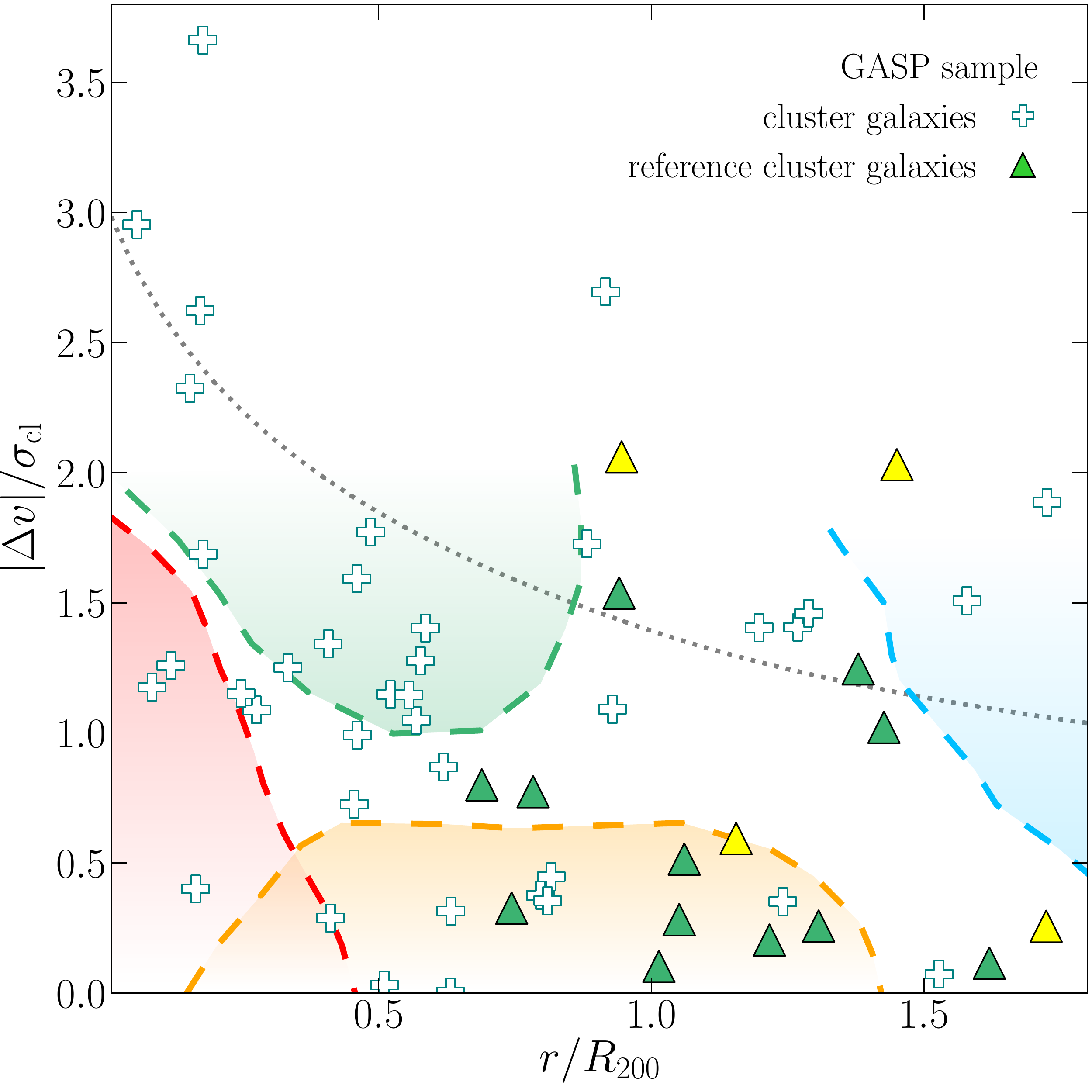}
\includegraphics[width=\columnwidth]{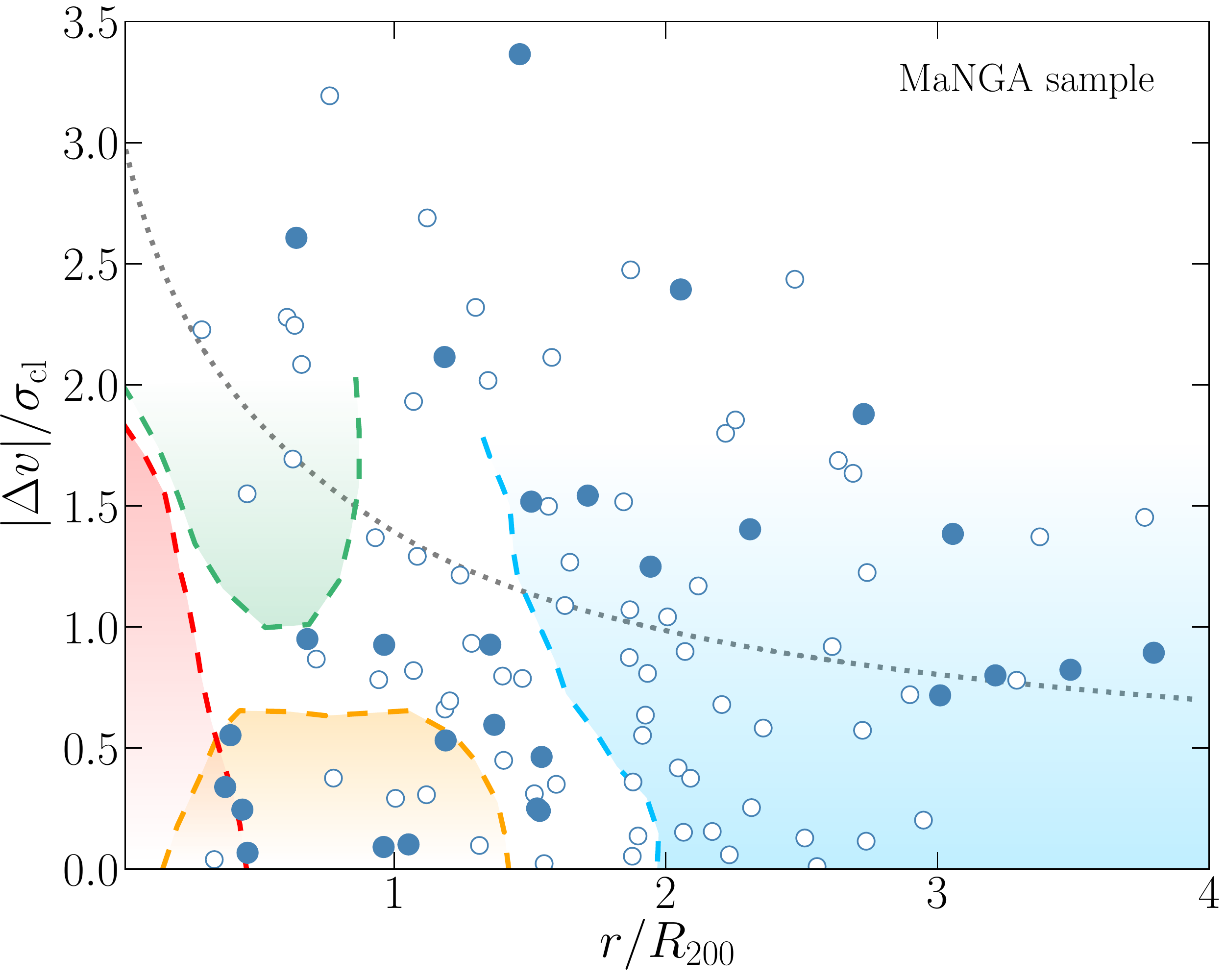}
\caption{Upper panel: phase space of cluster GASP galaxies. Filled triangles indicate reference galaxies, color-coded as in Fig.~\ref{fig:Mslope_OHslope} (yellow and green indicate galaxies closer and more distant, respectively, from the line that delimits the forbidden region -- Eq.~\ref{eq:fb}). Lower panel: phase space of cluster MaNGA galaxies. Filled dots indicate galaxies with metallicity gradients larger than the 75th percentile.
Blue, green, orange, and red shaded areas identify the regions of the diagram mainly populated by first, recent, intermediate, and ancient infallers in the cluster (see the text for details). Gray dotted curve shows the escape velocity in a \citet{navarro1996} halo assuming a concentration value of 6. \label{fig:PhaseSpace}}
\end{figure}

The RPS galaxies have most probably fallen only recently into the clusters ($<2$ Gyr ago) with radial orbits \citep{jaffe2018}, thus they have experienced the environmental effect only for a relatively short time (typically less than 1 Gyr). This could suggest that their chemical properties are more similar to those of field galaxies with respect to longstanding cluster members.
Reference cluster galaxies, instead, could travel on more circular orbits and have felt the influence of the cluster environment on their gas for longer times.
To investigate how long ago the reference cluster galaxies entered the cluster, in the top panel of Fig.~\ref{fig:PhaseSpace} we explore the phase-space diagram (galaxies velocity in the cluster versus the clustercentric distance) of the GASP cluster sample (see also \citealt{jaffe2018} and \citealt{gullieuszik2020}). Following \citet{rhee2017}, we highlight the regions where the majority of galaxies lie at a given epoch after they enter in the cluster halo, denoted as:
first (not fallen yet; blue), recent ($0<t_{\rm infall}<3.63$~Gyr; green), intermediate ($3.63<t_{\rm infall}<6.45$~Gyr; orange), and ancient ($6.45<t_{\rm infall}<13.7$~Gyr; red) infallers. Obviously, these numbers need to be taken with caution and each galaxy can spread out beyond the corresponding region anyway.
According to \citet{jaffe2018} and \citet{gullieuszik2020}, RPS galaxies are mostly located at high velocities and/or low clustercentric distances, i.e.\ they are preferentially in the regions mainly populated by first and recent infallers, as expected.
Instead, reference cluster galaxies have, on average, low velocity, and are located around the virial radius. They therefore tend to avoid the regions of both recent and ancient infallers, as most of them are placed in the region preferentially occupied by intermediate infallers.
Their position in the phase-space diagram supports the idea that these objects have experienced a slow cluster effect for a long time. In addition, as we depict the reference cluster galaxies by the same color-code used in Fig.~\ref{fig:Mslope_OHslope}, it is worth noticing that galaxies with the flattest metallicity profiles (green triangles) are mostly enclosed in the phase-space region dominated by intermediate infallers.
In contrast, galaxies with the highest velocities or with the largest clustercentric distances have steeper metallicity profiles for their given stellar mass (yellow triangles), suggesting that they may have fallen in the cluster more recently.

In the bottom panel of Fig.~\ref{fig:PhaseSpace} we show the phase-space diagram of MaNGA cluster galaxies. This sample includes galaxies at larger virial radii (up to 4~$R_{200}$) than GASP and most of them are located in the regions mainly populated by first infallers. In Fig.~\ref{fig:PhaseSpace}, we indicate with filled dots galaxies with shallow metallicity profiles (metallicity gradients greater than the 75th percentile in the stellar bins explored in Sec.~\ref{sec:OHresults}): we observe that half of these galaxies are concentrated near or inside the region corresponding to intermediate infallers, and thus it is plausible that their flatter metallicity profiles are due to the long permanence in the cluster. The other half, instead, spreads above the line of escape velocity of a \citet{navarro1996} halo (concentration value of 6), thus, we have not clues of their nature, even though they could have been pre-processed in groups or merging clusters before entering the main cluster.
The above analyses of cluster galaxies in GASP and MaNGA samples are, overall, in agreement, and we find promising hints of a connection between flat metallicity profile in cluster galaxies and infall time in the host halo.


\section{Conclusions}\label{sec:conclusion}

Radial gradients of gas metallicity provide important constrains to understand the formation and evolution of disk galaxies, and their study is attracting a wide attention.
In this paper, we have examined a sample of 85 disk galaxies from the GASP survey aiming to investigate the dependence of the ionized gas metallicity gradients on stellar mass and the environment. We also use a larger sample drawn from  MaNGA  which allows us to confirm the results obtained using the GASP sample on a more statistical ground and to investigate the role of the gas content in driving the metallicity gradients. 
The main results can be summarized as follows:
\begin{itemize}
    \item[i)] the great majority of the observed metallicity gradients are negative, meaning that the central regions are more metal-rich than the outer ones, in agreement with the inside-out formation of galaxies;
    \item[ii)] in both GASP and MANGA, the  metallicity profiles are almost flat at the lowest and the highest masses, while steep gradients and a larger scatter are observed at intermediate masses ($\sim10^{10.3}\,{\rm M_\odot}$);
    \item[iii)]  at least up to $10^{10.3}\,{\rm M_\odot}$, the GASP sample shows a clear offset between metallicity gradients of cluster and field galaxies, where the former present flatter profiles than the latter. This difference is also found in MaNGA galaxies at all stellar masses.
\end{itemize}

Thanks to GASP and MaNGA data, we are also able to study the stellar mass surface density profiles and compare them to the metallicity gradients. We find that:
\begin{itemize}
    \item[iv)] the stellar mass profile steepen with the global stellar mass, in agreement with previous works;
    \item[v)] according to the GASP sample, metallicity gradients mildly correlate with the stellar mass gradients ($PCC=0.28$); this correlation is the tightest for those galaxies that follow closely the metallicity gradient - stellar mass anti-correlation. Specifically, we do not observe galaxies with flat stellar mass profiles and steep metallicity profiles, while flat metallicity profile are found in galaxies with steep stellar mass profiles. The median trends of MaNGA galaxies are qualitatively in agreement with GASP, even if the MaNGA sample is characterized by a  larger scatter.
\end{itemize}
This last result
suggests
that metallicity profiles are initially shaped by in-situ star formation that enriched the ISM during the inside-out formation of galaxies, entailing the steepening of the metallicity profiles with the stellar mass. Additionally, mechanisms, mainly dominant at high stellar masses, are at work to flatten the radial distribution of the gas metallicity.

Thanks to MaNGA data, we are able to study the connection with the galaxy gas content. We find that:
\begin{itemize}
    \item[vi)] for any given stellar mass, the steepness of the metallicity profile correlates with the galaxy gas fraction, i.e., galaxies with flat metallicity gradients have lower gas fraction values than ones with steep gradients;
    \item[vii)] at the same time, as the gas fraction decreases and the metallicity profiles flatten, the metallicity at $R_{\rm e}$ increases.
\end{itemize}

The absence of low-mass galaxies with steep metallicity profiles, the correlation between gas metallicity gradients and the stellar mass surface density gradients, and the secondary dependence of the metallicity gradients on the global gas fraction provide  important constrains for chemical evolution models.

The dependence on the gas content might be at the origin of the differences between the metallicity profiles of field and cluster galaxies. Indeed, the latter could suffer by suppression of gas accretion according to their orbital trajectory and permanence inside the hostile cluster environment. Our work, carried out both with GASP and MaNGA, suggests that: \begin{itemize}
    \item[viii)] cluster galaxies with flatter metallicity profiles have presumably fallen in the host halo sooner than those with steeper profiles.
    Thus, they have experienced the cluster effect on the infalling gas for a long time.
    Pre-processing mechanisms can also be involved, and may be statistically identified from the analysis of the phase-space diagram;
    \item[ix)] metallicity gradients of galaxies undergoing ram-pressure stripping do not deviate much from the values found for field galaxies in the considered mass range.
\end{itemize}
Given the paucity of detailed studies on metallicity profiles of cluster galaxies, we advise additional dedicated works in order to shed light on this class of objects in comparison to field galaxies.
To conclude, our results of gas metallicity gradients in disk galaxies can be useful constraints for testing both cosmological simulations and analytical models of galaxy formation and chemical evolution.
\clearpage

\begin{acknowledgments}
We thank the rest of the GASP team for the support and the useful discussions. We thank the anonymous referee for their comments that have improved the paper.
Based on observations collected at the European Organization for Astronomical Research in the Southern Hemisphere under ESO programme 196.B-0578. This project has received funding from the European Research Council (ERC) under the European Union's Horizon 2020 research and innovation programme (grant agreement No. 833824, PI Poggianti). We acknowledge funding from the INAF main-stream funding programme (PI B.~Vulcani). B.~V.\ and M.~G.\ acknowledge the Italian PRIN-Miur 2017 (PI A.~Cimatti). J.~F.\ acknowledges financial support from the UNAM- DGAPA-PAPIIT IN111620 grant, M\'exico.

Funding for the Sloan Digital Sky Survey IV has been provided by the Alfred P. Sloan Foundation, the U.S. Department of Energy Office of Science, and the Participating Institutions. 
SDSS-IV acknowledges support and resources from the Center for High Performance Computing  at the University of Utah. The SDSS website is www.sdss.org.
SDSS-IV is managed by the Astrophysical Research Consortium for the Participating Institutions of the SDSS Collaboration including the Brazilian Participation Group, the Carnegie Institution for Science, Carnegie Mellon University, Center for Astrophysics | Harvard \& Smithsonian, the Chilean Participation Group, the French Participation Group, Instituto de Astrof\'isica de Canarias, The Johns Hopkins University, Kavli Institute for the Physics and Mathematics of the Universe (IPMU) / University of Tokyo, the Korean Participation Group, Lawrence Berkeley National Laboratory, Leibniz Institut f\"ur Astrophysik Potsdam (AIP),  Max-Planck-Institut f\"ur Astronomie (MPIA Heidelberg), Max-Planck-Institut f\"ur Astrophysik (MPA Garching), Max-Planck-Institut f\"ur Extraterrestrische Physik (MPE), National Astronomical Observatories of China, New Mexico State University, New York University, University of Notre Dame, Observat\'ario Nacional / MCTI, The Ohio State University, Pennsylvania State University, Shanghai Astronomical Observatory, United Kingdom Participation Group, Universidad Nacional Aut\'onoma de M\'exico, University of Arizona, University of Colorado Boulder, University of Oxford, University of Portsmouth, University of Utah, University of Virginia, University of Washington, University of Wisconsin, Vanderbilt University, and Yale University. The MaNGA data used in this work is publicly
available at \url{http://www.sdss.org/dr15/manga/manga-data/}.

\end{acknowledgments}

\software{MaNGA data reduction pipeline \citep{law2016}, MUSE pipeline \citep{bacon2010}, MaNGA data analysis pipeline (v2.2.1; \citealt{westfall2019}; \citealt{belfiore2019b}), {\sc sinopsis} \citep{fritz2017}, {\sc kubeviz} \citep{fossati2016}, {\sc pPXF} \citep{cappellari2004}, {\sc pyqz} \citep{dopita2013,vogt2015}, {\sc emcee} \citep{foreman2013}, {\sc fit3d} \citep{sanchez2006,sanchez2016}}

\newpage
\appendix

\section{Stellar mass surface density profile}\label{app:Mprofile}

\begin{figure}[h!]
    \centering
    \includegraphics[width=0.6\columnwidth]{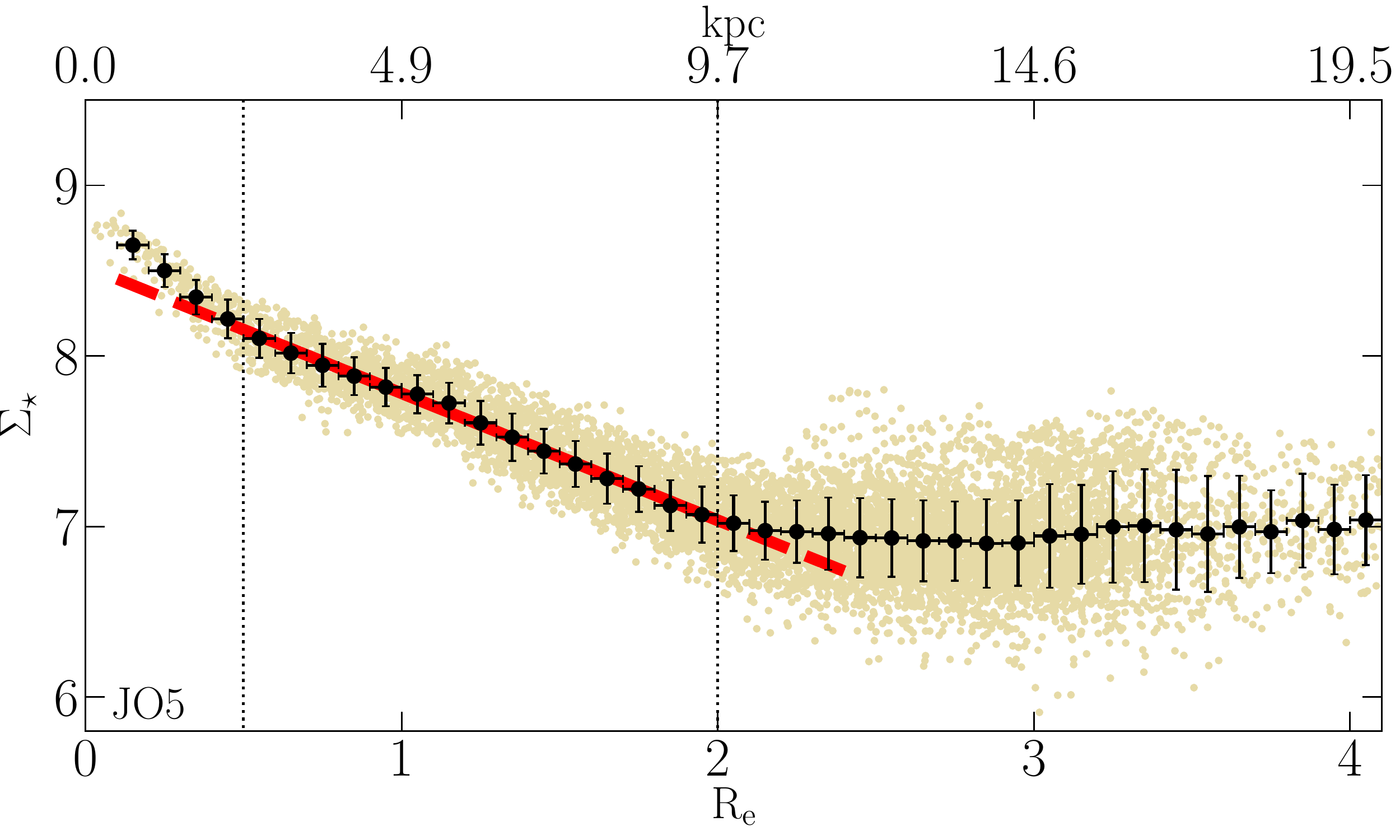}
    \caption{Stellar mass surface density profile of JO5. Small points correspond to the value of each single spaxel, while the black dots are the azimuthal mean in radial bins of $0.1\,{\rm R_e}$. The red line indicates the best linear fit between 0.5 and $2\,{\rm R_e}$.}
    \label{fig:Mprofile}
\end{figure}

Here, we provide an example of the typical stellar mass surface density profile of a GASP galaxy, that we fitted to derive the radial gradient. In Fig.~\ref{fig:Mprofile} we show the profile of JO5. The stellar mass radially decreases up to $2\,{\rm R_e}$ with a linear trend, while at higher radii the profile becomes flat. Within $0.5\,{\rm R_e}$, we observe a slight steepening presumably due to the presence of the bulge. All GASP galaxies have a similar stellar mass profile, where the radius of the outer plateau (if present) can vary between 2 and $4\,{\rm R_e}$. Given this common shape of the profiles, we are confident to obtain a robust estimate of the stellar mass gradient by a simple linear fitting between 0.5 and $2\,{\rm R_e}$.

\section{Metallicity and stellar mass gradient values}\label{app:gradients}

\startlongtable
\begin{deluxetable*}{lcccccc}
\tablecaption{Properties of the GASP galaxies. Columns are: (1) GASP ID number; (2) and (3) Equatorial coordinates of the galaxy center; (4) environment (F=field, C=cluster); (5) logarithm of the stellar mass; (6) gas metallicity gradient; (7) stellar mass surface density gradient. \label{tab:tabella}}
\tablehead{\colhead{ID} & \colhead{R.A.} & \colhead{Decl.} & \colhead{env} & \colhead{$\log(M_\star/M_\odot)$} & \colhead{$\alpha_{\rm O/H}$} & \colhead{$\alpha_{M\star}$}\\
\colhead{} & \colhead{(J2000)} & \colhead{(J2000)} & \colhead{} & \colhead{} & \colhead{$({\rm dex\,R_e^{-1}})$} & \colhead{$({\rm dex\,R_e^{-1}})$}\\
\colhead{(1)} & \colhead{(2)} & \colhead{(3)} & \colhead{(4)} & \colhead{(5)} & \colhead{(6)} & \colhead{(7)}}
\startdata
JO73 & 22:04:25.99 & $-$05:14:47.041 & F & $10.0_{-0.1}^{+0.1}$ & $-0.43\pm0.08$ & $-0.58\pm0.09$ \\
P11695 & 10:46:14.89 & $+$00:03:00.838 & F & $10.10_{-0.08}^{+0.07}$ & $-0.22\pm0.03$ & $-0.76\pm0.09$ \\
P12823 & 10:52:24.04 & $-$00:06:09.882 & F & $10.3_{-0.1}^{+0.1}$ & $-0.082\pm0.007$ & $-0.7\pm0.1$ \\
P13384 & 10:53:03.15 & $-$00:13:30.932 & F & $9.9_{-0.1}^{+0.1}$ & $-0.32\pm0.06$ & $-0.74\pm0.07$ \\
P14672 & 11:01:55.10 & $+$00:11:41.095 & F & $9.90_{-0.09}^{+0.07}$ & $-0.3\pm0.1$ & $-0.67\pm0.07$ \\
P15703 & 11:06:33.28 & $+$00:16:48.192 & F & $11.00_{-0.09}^{+0.07}$ & $-0.09\pm0.02$ & $-0.72\pm0.04$ \\
P17048 & 11:12:38.27 & $+$00:08:01.150 & F & $9.6_{-0.1}^{+0.1}$ & $-0.24\pm0.04$ & $-0.64\pm0.07$ \\
P17945 & 11:15:26.45 & $+$00:16:11.586 & F & $9.7_{-0.1}^{+0.1}$ & $-0.26\pm0.04$ & $-0.64\pm0.08$ \\
P18060 & 11:14:59.28 & $-$00:00:43.199 & F & $9.1_{-0.1}^{+0.1}$ & $-0.12\pm0.09$ & $-0.63\pm0.08$ \\
P19482 & 11:22:31.25 & $-$00:01:01.601 & F & $10.33_{-0.09}^{+0.07}$ & $-0.20\pm0.02$ & $-0.83\pm0.07$ \\
P20159 & 11:24:22.26 & $-$00:16:34.268 & F & $9.6_{-0.1}^{+0.1}$ & $-0.07\pm0.02$ & $-0.57\pm0.09$ \\
P20769 & 11:27:17.60 & $+$00:11:24.388 & F & $9.4_{-0.1}^{+0.1}$ & $-0.12\pm0.02$ & $-0.70\pm0.08$ \\
P20883 & 11:27:45.41 & $-$00:07:16.580 & F & $9.9_{-0.1}^{+0.1}$ & $-0.26\pm0.02$ & $-0.60\pm0.08$ \\
P21734 & 11:31:07.90 & $-$00:08:07.914 & F & $10.8_{-0.1}^{+0.1}$ & $-0.52\pm0.07$ & $-0.87\pm0.08$ \\
P25500 & 11:51:36.28 & $+$00:00:01.929 & F & $10.8_{-0.1}^{+0.1}$ & $-0.23\pm0.02$ & $-0.71\pm0.09$ \\
P40457 & 13:01:33.06 & $-$00:04:51.121 & F & $9.7_{-0.1}^{+0.1}$ & $-0.24\pm0.04$ & $-0.6\pm0.1$ \\
P42932 & 13:10:44.71 & $+$00:01:55.540 & F & $10.51_{-0.08}^{+0.07}$ & $-0.2\pm0.1$ & $-0.70\pm0.06$ \\
P45479 & 13:23:34.73 & $-$00:07:51.673 & F & $10.57_{-0.08}^{+0.07}$ & $-0.20\pm0.04$ & $-0.78\pm0.06$ \\
P48157 & 13:36:01.59 & $+$00:15:44.696 & F & $10.59_{-0.09}^{+0.08}$ & $-0.22\pm0.02$ & $-0.66\pm0.08$ \\
P5215 & 10:16:58.24 & $-$00:14:52.876 & F & $10.52_{-0.09}^{+0.07}$ & $-0.31\pm0.06$ & $-0.70\pm0.06$ \\
P57486 & 14:11:34.45 & $+$00:09:58.293 & F & $9.9_{-0.1}^{+0.1}$ & $-0.24\pm0.02$ & $-0.62\pm0.08$ \\
P59597 & 14:17:41.13 & $-$00:08:39.351 & F & $9.7_{-0.1}^{+0.1}$ & $-0.2\pm0.1$ & $-0.6\pm0.1$ \\
P63661 & 14:32:21.78 & $+$00:10:41.428 & F & $10.3_{-0.1}^{+0.1}$ & $-0.29\pm0.05$ & $-0.75\pm0.08$ \\
P648 & 10:01:27.74 & $+$00:09:18.372 & F & $10.4_{-0.1}^{+0.1}$ & $-0.25\pm0.02$ & $-0.60\pm0.07$ \\
P669 & 10:02:00.62 & $+$00:10:44.299 & F & $10.49_{-0.08}^{+0.07}$ & $-0.14\pm0.02$ & $-0.75\pm0.06$ \\
P8721 & 10:34:08.65 & $+$00:00:03.189 & F & $10.72_{-0.07}^{+0.06}$ & $-0.22\pm0.02$ & $-0.78\pm0.08$ \\
P95080 & 13:12:08.75 & $-$00:14:20.334 & F & $10.1_{-0.1}^{+0.1}$ & $-0.33\pm0.03$ & $-0.69\pm0.09$ \\
P954 & 10:02:03.33 & $-$00:12:49.836 & F & $9.6_{-0.1}^{+0.1}$ & $-0.22\pm0.03$ & $-0.56\pm0.09$ \\
P96244 & 14:18:35.47 & $+$00:09:27.828 & F & $10.8_{-0.2}^{+0.1}$ & $-0.32\pm0.06$ & $-0.72\pm0.07$ \\
A3128\_B\_0148 & 03:27:31.09 & $-$52:59:07.655 & C & $9.8_{-0.1}^{+0.1}$ & $-0.12\pm0.02$ & $-0.8\pm0.1$ \\
A3266\_B\_0257 & 04:27:52.58 & $-$60:54:11.565 & C & $9.9_{-0.1}^{+0.1}$ & $-0.16\pm0.02$ & $-0.62\pm0.09$ \\
A3376\_B\_0261 & 06:00:13.68 & $-$39:34:49.232 & C & $10.53_{-0.09}^{+0.08}$ & $-0.25\pm0.02$ & $-0.84\pm0.06$ \\
A970\_B\_0338 & 10:19:01.65 & $-$10:10:36.924 & C & $10.06_{-0.09}^{+0.07}$ & $-0.12\pm0.02$ & $-0.65\pm0.09$ \\
JO102 & 03:29:04.69 & $-$52:50:05.364 & C & $10.01_{-0.09}^{+0.07}$ & $-0.08\pm0.02$ & $-0.6\pm0.1$ \\
JO112 & 03:40:06.02 & $-$54:02:27.300 & C & $9.61_{-0.08}^{+0.07}$ & $-0.22\pm0.08$ & $-0.40\pm0.08$ \\
JO113 & 03:41:49.17 & $-$53:24:13.680 & C & $9.7_{-0.1}^{+0.1}$ & $-0.22\pm0.08$ & $-0.8\pm0.1$ \\
JO119 & 06:29:59.10 & $-$54:47:38.690 & C & $10.4_{-0.1}^{+0.1}$ & $-0.10\pm0.05$ & $-0.7\pm0.1$ \\
JO123 & 12:53:01.03 & $-$28:36:52.584 & C & $9.9_{-0.1}^{+0.1}$ & $-0.18\pm0.02$ & $-0.5\pm0.1$ \\
JO128 & 12:54:56.84 & $-$29:50:11.184 & C & $9.9_{-0.1}^{+0.1}$ & $-0.4\pm0.1$ & $-0.6\pm0.1$ \\
JO13 & 00:55:39.68 & $-$00:52:35.981 & C & $9.82_{-0.08}^{+0.07}$ & $-0.30\pm0.04$ & $-0.61\pm0.08$ \\
JO135 & 12:57:04.30 & $-$30:22:30.313 & C & $10.99_{-0.08}^{+0.07}$ & $-0.11\pm0.02$ & $-0.7\pm0.1$ \\
JO138 & 12:56:58.51 & $-$30:06:06.284 & C & $9.7_{-0.2}^{+0.1}$ & $-0.09\pm0.03$ & $-0.47\pm0.09$ \\
JO141 & 12:58:38.38 & $-$30:47:32.200 & C & $10.7_{-0.2}^{+0.1}$ & $-0.23\pm0.04$ & $-0.71\pm0.08$ \\
JO144 & 13:24:32.43 & $-$31:06:59.036 & C & $10.5_{-0.2}^{+0.1}$ & $-0.12\pm0.01$ & $-0.7\pm0.1$ \\
JO147 & 13:26:49.73 & $-$31:23:45.511 & C & $11.03_{-0.09}^{+0.07}$ & $-0.030\pm0.004$ & $-0.6\pm0.1$ \\
JO149 & 13:28:10.53 & $-$31:09:50.200 & C & $8.8_{-0.2}^{+0.1}$ & $0.01\pm0.03$ & $-0.1\pm0.1$ \\
JO153 & 13:28:15.15 & $-$31:01:57.859 & C & $9.4_{-0.1}^{+0.1}$ & $-0.07\pm0.02$ & $-0.5\pm0.1$ \\
JO157 & 13:28:18.18 & $-$31:48:18.789 & C & $10.1_{-0.1}^{+0.1}$ & $-0.19\pm0.02$ & $-0.6\pm0.1$ \\
JO159 & 13:26:35.70 & $-$30:59:36.920 & C & $9.8_{-0.1}^{+0.1}$ & $-0.03\pm0.03$ & $-0.71\pm0.08$ \\
JO160 & 13:29:28.62 & $-$31:39:25.288 & C & $10.1_{-0.1}^{+0.1}$ & $-0.17\pm0.08$ & $-0.76\pm0.09$ \\
JO162 & 13:31:29.92 & $-$33:03:19.576 & C & $9.4_{-0.1}^{+0.1}$ & $-0.10\pm0.01$ & $-0.62\pm0.09$ \\
JO17 & 01:08:35.33 & $+$01:56:37.043 & C & $10.16_{-0.09}^{+0.08}$ & $-0.21\pm0.02$ & $-0.72\pm0.07$ \\
JO171 & 20:10:14.70 & $-$56:38:30.561 & C & $10.61_{-0.07}^{+0.06}$ & $-0.08\pm0.02$ & $-0.3\pm0.1$ \\
JO175 & 20:51:17.60 & $-$52:49:21.825 & C & $10.50_{-0.08}^{+0.06}$ & $-0.21\pm0.02$ & $-0.95\pm0.06$ \\
JO179 & 21:47:07.07 & $-$43:42:18.221 & C & $9.5_{-0.1}^{+0.1}$ & $-0.17\pm0.03$ & $-0.6\pm0.1$ \\
JO180 & 21:45:15.00 & $-$44:00:31.188 & C & $10.0_{-0.1}^{+0.1}$ & $-0.18\pm0.02$ & $-0.65\pm0.07$ \\
JO181 & 22:28:03.80 & $-$30:18:03.812 & C & $9.1_{-0.1}^{+0.1}$ & $-0.0\pm0.1$ & $-0.4\pm0.1$ \\
JO194 & 23:57:00.68 & $-$34:40:50.117 & C & $11.2_{-0.1}^{+0.1}$ & $-0.045\pm0.009$ & $-0.83\pm0.07$ \\
JO197 & 09:06:32.58 & $-$09:31:27.282 & C & $10.0_{-0.1}^{+0.1}$ & $-0.25\pm0.02$ & $-0.68\pm0.07$ \\
JO200 & 00:42:05.03 & $-$09:32:03.841 & C & $10.82_{-0.08}^{+0.06}$ & $-0.37\pm0.06$ & $-0.81\pm0.06$ \\
JO201 & 00:41:30.29 & $-$09:15:45.900 & C & $10.79_{-0.06}^{+0.05}$ & $-0.05\pm0.01$ & $-0.90\pm0.08$ \\
JO204 & 10:13:46.83 & $-$00:54:51.056 & C & $10.61_{-0.07}^{+0.06}$ & $-0.042\pm0.006$ & $-0.75\pm0.08$ \\
JO205 & 21:13:46.12 & $+$02:14:20.355 & C & $9.5_{-0.1}^{+0.1}$ & $-0.2\pm0.1$ & $-0.57\pm0.09$ \\
JO206 & 21:13:47.41 & $+$02:28:34.383 & C & $10.96_{-0.05}^{+0.04}$ & $-0.10\pm0.02$ & $-0.9\pm0.1$ \\
JO27 & 01:10:48.56 & $-$15:04:41.611 & C & $9.5_{-0.1}^{+0.1}$ & $-0.06\pm0.02$ & $-0.6\pm0.1$ \\
JO28 & 01:10:09.31 & $-$15:34:24.507 & C & $9.36_{-0.08}^{+0.07}$ & $-0.21\pm0.04$ & $-0.47\pm0.08$ \\
JO41 & 12:53:54.79 & $-$15:47:20.096 & C & $10.20_{-0.07}^{+0.06}$ & $-0.08\pm0.03$ & $-0.74\pm0.08$ \\
JO45 & 01:13:16.58 & $+$00:12:05.839 & C & $9.2_{-0.1}^{+0.1}$ & $-0.06\pm0.03$ & $-0.63\pm0.08$ \\
JO47 & 01:15:57.67 & $+$00:41:35.938 & C & $9.60_{-0.07}^{+0.06}$ & $-0.09\pm0.03$ & $-0.43\pm0.09$ \\
JO49 & 01:14:43.85 & $+$00:17:10.091 & C & $10.68_{-0.06}^{+0.06}$ & $-0.098\pm0.009$ & $-0.80\pm0.08$ \\
JO5 & 10:41:20.38 & $-$08:53:45.559 & C & $10.3_{-0.1}^{+0.1}$ & $-0.33\pm0.03$ & $-0.74\pm0.08$ \\
JO60 & 14:53:51.57 & $+$18:39:06.364 & C & $10.4_{-0.1}^{+0.1}$ & $-0.22\pm0.06$ & $-0.6\pm0.1$ \\
JO68 & 21:56:22.00 & $-$07:54:28.971 & C & $10.0_{-0.1}^{+0.1}$ & $-0.21\pm0.02$ & $-0.73\pm0.09$ \\
JO69 & 21:57:19.20 & $-$07:46:43.794 & C & $9.9_{-0.1}^{+0.1}$ & $-0.3\pm0.1$ & $-0.5\pm0.1$ \\
JO70 & 21:56:04.07 & $-$07:19:38.020 & C & $10.5_{-0.1}^{+0.1}$ & $-0.34\pm0.05$ & $-0.95\pm0.08$ \\
JO85 & 23:24:31.36 & $+$16:52:05.340 & C & $10.7_{-0.1}^{+0.1}$ & $-0.21\pm0.03$ & $-0.7\pm0.1$ \\
JO89 & 23:26:00.60 & $+$14:18:26.291 & C & $9.72_{-0.09}^{+0.07}$ & $-0.16\pm0.04$ & $-0.74\pm0.07$ \\
JO93 & 23:23:11.74 & $+$14:54:05.013 & C & $10.54_{-0.08}^{+0.07}$ & $-0.34\pm0.02$ & $-0.86\pm0.08$ \\
JO95 & 23:44:26.66 & $+$09:06:55.839 & C & $9.37_{-0.09}^{+0.08}$ & $-0.14\pm0.03$ & $-0.4\pm0.1$ \\
JW10 & 04:39:18.19 & $-$21:57:49.627 & C & $10.0_{-0.1}^{+0.1}$ & $-0.15\pm0.01$ & $-0.6\pm0.1$ \\
JW100 & 23:36:25.06 & $+$21:09:02.529 & C & $11.5_{-0.1}^{+0.1}$ & $-0.04\pm0.01$ & $-0.68\pm0.08$ \\
JW115 & 12:00:47.95 & $-$31:13:41.635 & C & $9.7_{-0.1}^{+0.1}$ & $-0.10\pm0.03$ & $-0.6\pm0.1$ \\
JW29 & 12:57:49.48 & $-$17:39:57.095 & C & $9.50_{-0.09}^{+0.07}$ & $0.01\pm0.06$ & $-0.50\pm0.08$ \\
JW39 & 13:04:07.71 & $+$19:12:38.486 & C & $11.21_{-0.08}^{+0.07}$ & $-0.04\pm0.01$ & $-0.76\pm0.08$ \\
JW56 & 13:27:03.03 & $-$27:12:58.205 & C & $9.1_{-0.1}^{+0.1}$ & $-0.08\pm0.06$ & $-0.6\pm0.2$ \\
\enddata
\end{deluxetable*}
\onecolumngrid

\section{Best-fit parameters}\label{app:bestparam}
\hfill\vspace{-12pt}
\begin{deluxetable*}{lcccccccc}[h!]
\tablecaption{Best-fit parameters. Columns are: (1) GASP ID number; (2) best-fit profile function (sl=`single-linear', dl=`double-linear', tl=`triple-linear'; see Sec.~\ref{sec:method_OHgradients} for detail); (3) gas metallicity gradient; (4) gas metallicity at $R_{\rm e}$; (5) break radius of the `double-linear' profile; (6) and (7) inner and outer radii of the `triple-linear' profile, respectively; (8) and (9) inner and outer gas metallicity gradient, respectively. \label{tab:tabella2}}
\tablehead{\colhead{ID} & \colhead{function} & \colhead{$\alpha_{\rm O/H}$} & \colhead{$Z_{R\rm e}$} & \colhead{$R_{\rm b}$} & \colhead{$R_{\rm in}$} & \colhead{$R_{\rm out}$} & \colhead{$\alpha_{\rm in}$} & \colhead{$\alpha_{\rm out}$}\\
\colhead{} & \colhead{} & \colhead{$({\rm dex\,R_e^{-1}})$} & \colhead{} & \colhead{($R_{\rm e}$)} & \colhead{($R_{\rm e}$)} & \colhead{($R_{\rm e}$)} & \colhead{$({\rm dex\,R_e^{-1}})$} & \colhead{$({\rm dex\,R_e^{-1}})$}\\
\colhead{(1)} & \colhead{(2)} & \colhead{(3)} & \colhead{(4)} & \colhead{(5)} & \colhead{(6)} & \colhead{(7)} & \colhead{(8)} & \colhead{(9)}}
\startdata
JO73 & dl & $-0.43\pm0.08$ & $8.57\pm0.04$ & $1.1\pm0.4$ & \dots & \dots & \dots & $-0.1\pm0.2$\\
P11695 & dl & $-0.22\pm0.03$ & $8.71\pm0.02$ & $1.5\pm0.4$ & \dots & \dots & \dots & $-0.03\pm0.06$\\
P12823 & sl & $-0.082\pm0.007$ & $9.112\pm0.009$ & \dots & \dots & \dots & \dots & \dots\\
P14672 & dl & $-0.3\pm0.1$ & $8.96\pm0.03$ & $1.2\pm0.4$ & \dots & \dots & $-0.07\pm0.05$ & \dots\\
P15703 & dl & $-0.09\pm0.02$ & $9.16\pm0.02$ & $1.3\pm0.3$ & \dots & \dots & $0.1\pm0.1$ & \dots\\
P18060 & sl & $-0.12\pm0.09$ & $8.39\pm0.04$ & \dots & \dots & \dots & \dots & \dots\\
P21734 & tl & $-0.52\pm0.07$ & $8.97\pm0.02$ & \dots & $0.6\pm0.1$ & $1.8\pm0.3$ & $-0.05\pm0.06$ & $-0.1\pm0.1$\\
\enddata
\tablecomments{Table~\ref{tab:tabella2} is published in its entirety in the machine-readable format. A portion is shown here for guidance regarding its form and content.}
\end{deluxetable*}
\onecolumngrid

\section{Stellar mass-gas fraction-gas metallicity gradient surface}\label{app:surfaces}
\begin{figure*}[h!]
    \centering
    \includegraphics[width=0.49\textwidth]{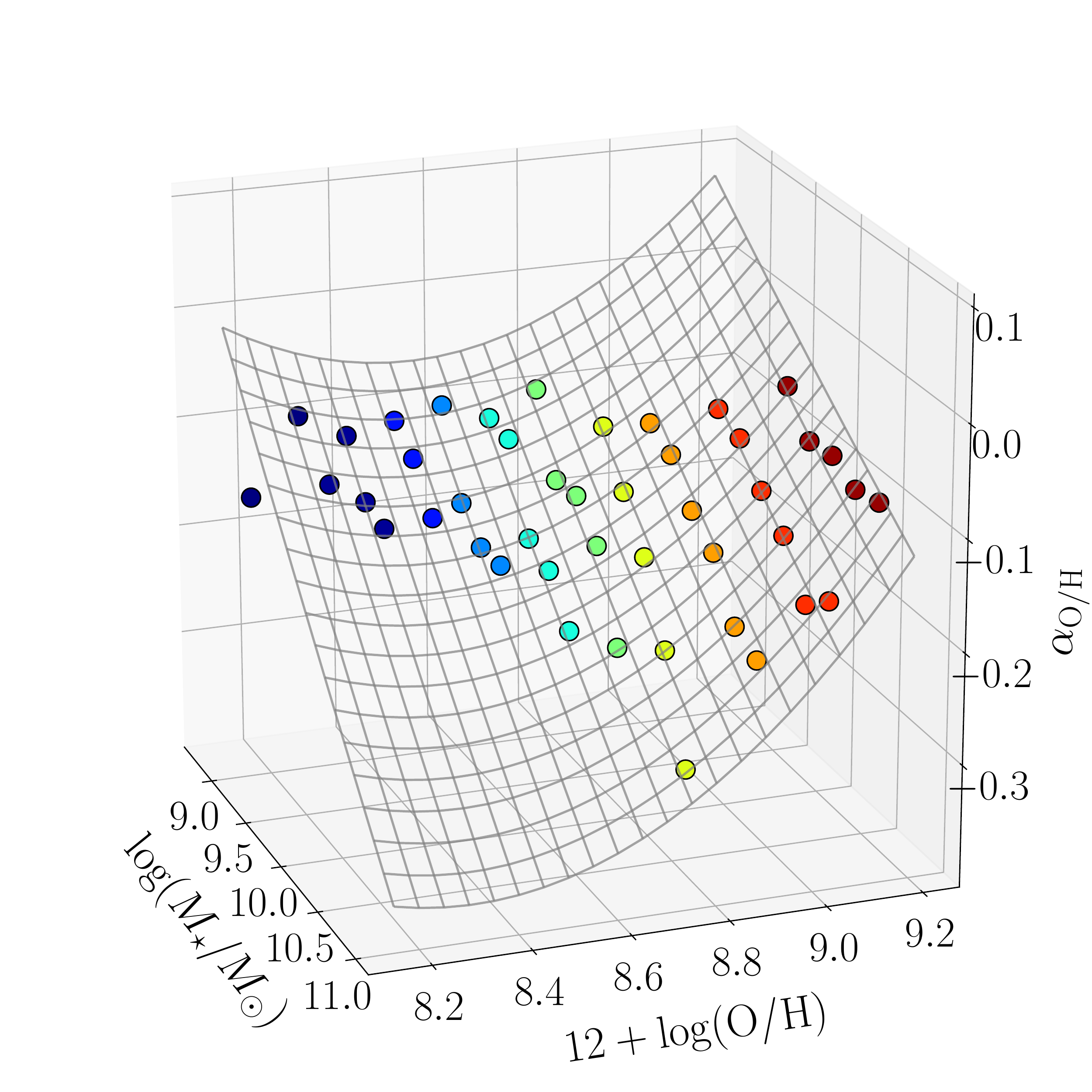}
    \includegraphics[width=0.49\textwidth]{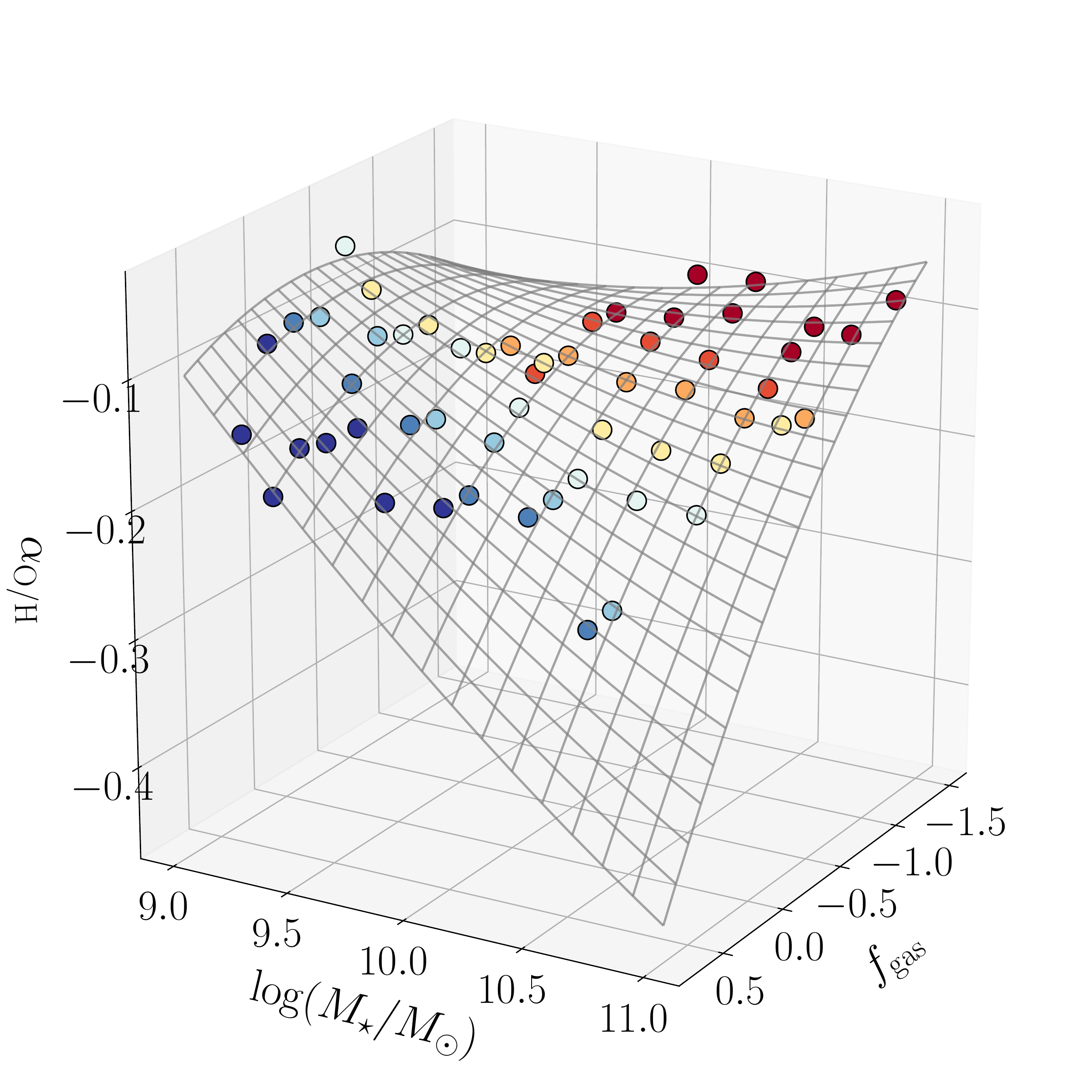}
    \caption{Left panel: Distribution of the MaNGA gas metallicity gradients as a function of stellar mass and gas metallicity at $R_{\rm e}$. Dots are the mean metallicity gradients in bins of the other two variables, color-coded by the gas metallicity, as a guide. The surface is the 2nd-order polynomial fit described by Eq.~\ref{eq:aOH_M_OHre}. Right panel: Distribution of the MaNGA gas metallicity gradients as a function of stellar mass and gas fraction. Dots are the mean metallicity gradients in bins of the other two variables, color-coded by the gas fraction, as a guide. The surface is the 2nd-order polynomial fit described by Eq.~\ref{eq:aOH_M_fgas}.}
    \label{fig:surfaces}
\end{figure*}


\end{document}